\documentclass[11pt]{iopart}

\usepackage{amsmath}
\usepackage{amsgen}
\usepackage{amssymb}
\usepackage{setspace}
\usepackage{setstack}
\usepackage{cite}
\usepackage{color}
\usepackage{yfonts}
\usepackage{graphics}
\usepackage{graphicx}
\usepackage{eufrak}
\usepackage[colorlinks=true, allcolors=blue]{hyperref}

\begin{document}

\title[]{Constraining the Cosmological Evolution of Post-Newtonian Parameters with Gravitational Wave Signals from Compact Binary Inspirals}

\author{Oliver Pitt$^{1}$ and Timothy Clifton$^{2}$}
\address{Department of Physics \& Astronomy, Queen Mary University of London, UK.}
\ead{$^1$o.pitt@qmul.ac.uk, $^2$t.clifton@qmul.ac.uk}

\vspace{0.5cm}
\begin{abstract}
Gravitational waves from compact binary inspirals offer a new opportunity to constrain the cosmological time dependence of gravitational coupling parameters, due to the high precision of the observations themselves as well as the significant cosmological redshifts at which such systems exist. We calculate theory-independent equations of motion for compact objects in a binary system, implementing a new approach to sensitivities, and subsequently determine the gravitational wave signal that one should expect to measure from their inspiral. Expressions for the wave phase and amplitude are derived in terms of post-Newtonian gravitational coupling parameters, radiative flux parameters, and compact body sensitivities. These results complement recent attempts to gain theory-independent constraints on the time-evolution of gravitational coupling parameters from cosmological probes, and represent a new opportunity to constrain modified gravity with gravitational wave data.
\end{abstract}

\vspace{0.5cm}
\section{Introduction}

Einstein's theory of General Relativity (GR), one of the cornerstones of twentieth century physics, has been extensively tested in the Solar System and other nearby astrophysical systems \cite{will2014confrontation}. The established method for such tests is to postulate a set of theory-independent coupling parameters between matter and gravitational potentials, and then to constrain the allowed values of these parameters with observations. The industry-standard approach for such tests uses the {\it Parameterized Post-Newtonian} (PPN) formalism, which is the foundation of virtually all tests of gravity in weak-field astrophysical systems \cite{will2018theory}.

A benefit of the PPN approach is that it allows tests of gravity to be performed without having to consider one theory at a time, which would be a particularly laborious task given the large number of such theories in the literature \cite{clifton2012modified}. It also has the added bonus of allowing one to directly identify which aspects of the gravitational interaction are constrained by which observables. As such, it functions as a particularly useful foil against which one can make precise constraints on possible deviations from the predictions of Einstein's equations, with constraints having a direct physical meaning.

Due to these successes, it is of considerable interest to extend the PPN formalism beyond its original application in isolated weak-field systems in the late Universe, and into new physical environments. Gravitational wave signals from the inspiral of compact astrophysical bodies present themselves as an ideal opportunity to do this, not only because they offer a new and highly relativistic environment to test gravity, but also because they can be observed from great distances. This means the gravitational waves could have been emitted a long time ago, which in some cases could be an order-one fraction of the age of the Universe. Such large separations in time and space allow cosmological variations of gravitational coupling parameters to be investigated and constrained, providing the possibility to extend studies of the cosmological time dependence of Newton's constant \cite{uzan2011varying} to the full suite of PPN parameters \cite{sanghai2017parameterized, clifton2019parametrizing, anton2022momentum, thomas2023scale, thomas2024constraining}.

To these ends, our goal is to characterize the phase and amplitude of gravitational wave signals during their early inspiral in a theory-independent manner using the PPN formalism as a starting point for the conservative two-body dynamics. As the objects in question are expected to be compact, with a strong gravitational field, we introduce a theory-independent concept of `sensitivity' (as familiar from scalar-tensor theories of gravity \cite{eardley1975observable}). This, together with a parametrized version of the energy lost due to gravitational radiation, is then sufficient to calculate the emitted gravitational wave signal. The observed detector strain is specified by this result, up to the effects of any modifications in the propagation equations between source and observer \cite{yunes2009fundamental}.
We expect our study to complement approaches that seek to parametrize the gravitational wave signal directly (see e.g. Ref.\cite{krishnendu2021testing}). Indeed, we find that we can express these phenomenological parametrizations in terms of the PPN gravitational coupling parameters, sensitivities, and luminosity parameters, in the spirit of Ref. \cite{sampson2013rosetta}. 

Our presentation will proceed as follows: In Section \ref{sec:theory} we consider the theoretical foundations of the gravitational physics we will deploy, including the parametrized post-Newtonian formalism, the idea of sensitivities, and a generalized luminosity equation. This is followed in Section \ref{sec:eom} by a theory-independent derivation of the equations of motion of compact bodies to first post-Newtonian order, including a new approach to sensitivities. In Section \ref{sec:gws} we calculate expressions for the gravitational wave phase and amplitude, and in Section \ref{sec:st} we then demonstrate that our approach can accommodate scalar-tensor theories of gravity. This is followed by a preliminary study of the implications of existing gravitational wave observations on the time-dependence of the PPN parameters in Section \ref{sec:constraints}, a discussion of these results in Section \ref{sec:dis}, and some concluding remarks in Section \ref{sec:conc}. Supplementary material is provided in appendices, and we work in units such that $G=c=1$ at the present time.

\section{Theoretical Foundations}
\label{sec:theory}

Our approach requires inputs from several different areas of gravitational physics, so in this section we will outline the necessary foundational concepts from each of these.

\subsection{Parametrized Post-Newtonian (PPN) Theory}

The PPN approach is based on the weak-field and slow-motion post-Newtonian expansion of Minkowski space. The order-of-smallness of this expansion is denoted by $\eta \sim v$, where $v \ll 1$ is the scalar 3-velocity of matter in the space-time\footnote{In most astrophysical systems, including the Solar System, $v \lesssim 10^{-2}$ (see e.g.  Ref. \cite{goldberg2017cosmology}).}. Assuming that all fields involved in the gravitational interaction vary on time-scales corresponding to this velocity, we are led to time derivatives adding an extra order-of-smallness to a quantity when compared to spatial derivatives, such that
\begin{equation}
    \frac{\frac{\partial}{\partial t}}{\frac{\partial}{\partial x}} \sim \eta \, .
    \label{placeholder}
\end{equation}
Using these orders of magnitude in the Euler equation immediately tells us that the leading-order part of the Newtonian potential is
\begin{equation}
    U \sim v^2 \sim \eta^2 \, ,
    \label{eq:placeholder}
\end{equation}
where $\nabla^2 U = - 4 \pi \, \rho^*$, and where $\rho^*=\rho + \frac{1}{2} v^2 +3 \gamma U$ is the `conserved density'. When considering the equations of motion of massive bodies the Newtonian order (0PN) occurs at $\mathcal{O}(\eta^2)$, the post-Newtonian order (1PN) occurs at $\mathcal{O}(\eta^4)$, with higher post-Newtonian orders occur at higher even-orders of $\eta$. For compact binaries, the post-Newtonian approximation is expected to be valid for the early stages of the inspiral.

In standard post-Newtonian gauge, and for point masses, semi-conservative theories have the following metric components in the PPN approach \cite{will1972conservation}:
\begin{align}
   \hspace{-0.5cm} g_{00} &= -1 +2 \alpha U - 2 \beta U^2 +(2 \alpha^2 -4 \beta ) \sum_{a,b} \frac{m_{{\rm G}a} m_{{\rm G} b}}{r_a r_{ab}} + (2 \gamma + \alpha) \sum_a \frac{m_{{\rm G} a} v_a^2}{r_a} +\mathcal{O}(\eta^5) \label{ppn1} \\
    g_{0i} &= - \frac{1}{2} (4 \gamma +4 \alpha + \alpha_1 ) \sum_a \frac{m_{{\rm G}a} v_a^i}{r_a} -\frac{1}{2} (\alpha +\alpha_2 ) \sum_a \frac{m_{{\rm G}a}}{r_a^3} \, (\vec{v}_a \cdot \vec{r}_a) \, r_a^i +\mathcal{O}(\eta^4) \label{ppn2} \\[4pt]
    g_{ij} &= (1+2 \gamma U) \delta_{ij} +\mathcal{O}(\eta^3)\, , \label{ppn3}
\end{align}
where $U=\sum_a m_{{\rm G}a}/r_a$ is the Newtonian gravitational potential and $\vec{r}_{ab} \equiv \vec{r}_{a}-\vec{r}_b$. Here we have included $\alpha$ as described in Appendix A, neglected the Whitehead term \cite{lee1974conservation}, and denoted the active gravitational mass by $m_{\rm G}$. This form of the metric is specified as an ansatz, and assumed to be valid for all reasonable theories of gravity (theories that do not globally conserve energy and momentum, or that require Whitehead terms, can be included by adding terms \cite{will2018theory}).

The Newtonian potential in Equation (\ref{ppn1}) comes with a coefficient $\alpha$ corresponding to the Newtonian gravitational coupling strength, while the post-Newtonian potentials in (\ref{ppn1})-(\ref{ppn3}) are parametrised by the additional series of constants $\{\beta,\gamma, \alpha_1, \alpha_2\}$ that have their physical meanings given in Table \ref{tab:ppN values}. With the PPN test metric (\ref{ppn1})-(\ref{ppn3}) in hand, and assuming test particles follow geodesics, one is able to generate equations of motion for any number of gravitating point masses. In the following section we will generalize this approach so that it can also be applied to compact bodies, following the pioneering introduction of sensitivities by Eardley \cite{eardley1975observable}.

\begin{table}
    \centering
    \begin{tabular}{ccccc}
        Parameter& Physical meaning &  &  & \\
        \hline
         $\alpha$ & {\it Newtonian gravitational coupling strength} &  &  & \\
          $\beta$ & {\it The degree of non-linearity in the superposition of gravity} &  &  & \\
        $\gamma$ & {\it How much spatial curvature a unit mass produces}  &  &  & \\
        $\alpha_i$ & {\it The amplitude of preferred frame effects} &  &  & \
    \end{tabular}
    \caption{The PPN parameters for semi-conservative theories of gravity with global conservation laws for energy and momentum, together with their physical meanings.
    }
    \label{tab:ppN values}
\end{table}

\subsection{Sensitivities of Compact Bodies}

In most theories of gravity the Strong Equivalence Principle (SEP) is violated, and the centre-of-mass of self-gravitating bodies does not follow a geodesic of the space-time \cite{will2018theory}. At 1PN this violation is called `the Nordtvedt effect' \cite{nordtvedt1968equivalence}, which vanishes in Einstein's theory, but is otherwise expected to be non-zero. More generally, the world-lines of compact bodies (such as black holes and neutron stars) also deviate from time-like geodesics due to their interiors being poorly modelled by post-Newtonian methods. A new approach is therefore required, which is the reason for introducing `sensitivities'.

The key insight by Eardley is to recognise that extended bodies can be treated as point-like particles with environmentally-dependent masses \cite{eardley1975observable}. That is, a neutron star that has gravitational mass $m^{(0)}$ in Minkowski space can be modelled as a particle of mass $m \neq m^{(0)}$ when calculating its trajectory in the presence of other gravitating bodies. This difference in mass is to be understood as a difference in the binding energy of the star due to its local environment, which constitutes a violation of the strong equivalence principle, and which reduces to the Nordtvedt effect in the weak-field limit.

Eardley was considering scalar-tensor theories of gravity, and so naturally made the environmentally-dependent mass of his bodies a function of the value of the scalar field, $m=m(\phi)$. He showed, using matching procedures, that this reproduced the SEP-violating effects of compact bodies on their equations of motion. One can generalize Eardley's procedure by making the mass a function of arbitrary external fields $\psi_A$ \cite{Taherasghari:2022wfs}, which results in the following matter action:
\begin{equation}
\label{eq:eardley equation}
    I_m = - \sum_a \int m_a \left( \psi_A [x_a(\tau_a)] \right) d\tau_a \,
\end{equation}
where $x^i_a(\tau_a)$ parametrizes the world-line of the $a$th body in terms of its proper time $\tau_a$. 

Adding an appropriate gravitational action to Equation (\ref{eq:eardley equation}), and then varying this action with respect to the metric, $g_{\mu\nu}$, and the extra fields, $\psi_A$, results in equations of motion that are sensitive to the dependence of the mass upon the extra fields. Such sensitivity is expected to account for the effects of additional gravitational degrees of freedom in the matching region, and hence their influence on the internal structure of the body and therefore its motion. In the post-Newtonian limit one can expand $\psi_A$ around its background value, such that $\psi_A =\psi^{(0)}_A+\delta\psi_A$, which allows the effective mass of the point-particle representing the $a$th body to be written as
\begin{align}
    m_a(\psi_A) 
    &= m_{a}^{(0)} \left(1+ \sum_A s_a^A \, \delta \psi_A +\frac{1}{2} \sum_{A,B} s_a^{\prime \,AB} \, \delta \psi_A \, \delta \psi_B  \right) + \mathcal{O}(\delta \psi_A^3) \, ,
\end{align}
where $m_{a}^{(0)}=m_a(\psi_A^{(0)})$ and the sensitivities are defined by 
\begin{equation}
s_a^{A} \equiv \frac{\partial \ln m_a}{\partial \psi_A} \Bigg\vert_0\,
\qquad {\rm and } \qquad
s_a^{\prime \,AB} \equiv \frac{\partial^2 \ln m_a}{\partial \psi_A \, \partial \psi_B} \Bigg\vert_0\, .
\end{equation}
These parameters accommodate the strong-field effect of compact bodies within a post-Newtonian expansion of alternative gravity theories, and have been applied to vector-tensor theories \cite{foster2007strong, yagi2014constraints}, to higher-orders in scalar-tensor theories \cite{mirshekari2013compact, lang2014compact}, as well as to the modified Einstein-Infeld-Hofmann formalism \cite{Will_2018}. In Section \ref{sec:eom} we generalize these ideas to create a theory-independent concept of sensitivity.

\subsection{Theory-Independent Gravitational Luminosity}

Alternative theories of gravity typically excite not only quadrupolar gravitational radiation, but also dipolar and monopolar emission. Assuming the theory conserves energy globally, one expects terms that contain the following combinations of source multipoles to contribute to the rate of energy loss from a radiating system \cite{thorne1980multipole}:
\begin{equation} \label{mdi}
\left\{ \dot{\mathcal{M}}^2 \, , \, \ddot{\mathcal{D}}_i  \ddot{\mathcal{D}}^i\, , \, \dddot{\mathcal{I}}_{ij}  \dddot{\mathcal{I}}^{ij} \, , \, \dots  \right\} \, ,
\end{equation}
where $\mathcal{M} \sim \int \rho^* \, d^3x$ is the monopole, $\mathcal{D}^i \sim \int \rho^* \, x^i \, d^3x$ is the dipole, and $\mathcal{I}^{ij} \sim \int \rho^* \, x^ix^j d^3x$ is the quadrupole. The first two terms in  equation (\ref{mdi}) typically vanish in solutions of Einstein's equations, due to adherence to the strong equivalence principle. This is not the case in most alternative theories, however, where mass and linear momentum are not generically conserved at 1PN and above.

Aside from Einstein's theory, the luminosity equation for gravitational wave emission has only been analysed systematically and in detail for scalar-tensor \cite{damour1992tensor, mirshekari2013compact, bernard2022gravitational} and vector-tensor \cite{foster2007strong, yagi2014constraints} theories of gravity. In particular, there exists no expression for gravitational wave flux from a compact binary in the PPN approach, as the leading-order gravitational wave contribution is sourced only at 2PN and radiation-reaction terms in the equations of motion of massive particles typically occur at only 2.5PN. The standard approach used in the literature is therefore to add arbitrary parameters to the equation that governs energy loss from binary systems in Einstein's theory, with an additional term added to account for dipole emission \cite{will2018theory}: 
\begin{equation}
\label{eq:edot}
\dot{E} = -\frac{\mu^2 M^2}{r_{12}^4}\left[\frac{8}{15}(\kappa_1 \, v_{12}^2 - \kappa_2\, \dot{r}^2_{12}) +\kappa_3\,  v_{12}^4 +\frac{1}{3}\kappa_{\rm D}\, \mathcal{G}^2 \right] \, ,
\end{equation}
where $\kappa_1$, $\kappa_2$, $\kappa_3$ and $\kappa_{\rm D}$ are arbitrary parameters. In this equation $r_{12}$ is the orbital separation in a system with total mass $M=m_1+m_2$ and reduced mass $\mu= m_1 m_2/M$, where $v_{12}$ is the relative velocity. The first two terms in brackets reduce to the corresponding expressions for Einstein's theory when $\kappa_1=12$ and $\kappa_2=11$ \cite{peters1963gravitational}, while the third term corresponds to a post-Newtonian contribution. The last term in this equation corresponds to the energy loss in dipole radiation due to the difference in gravitational energy per unit mass,
\begin{equation} \label{curlG}
    \mathcal{G} = \frac{\Omega_1}{m_1} - \frac{\Omega_2}{m_2} \, , \qquad {\rm where} \qquad \Omega_i = -\frac{1}{2} \int_i \rho^* \, U \, d^3x \, .
\end{equation}
The form of these terms is motivated from the leading-order part of the multipole moments in binary systems \cite{thorne1980multipole}, and the expectation is that in alternative theories the dipole radiation should be sourced by the dipole moment of gravitational energy. Equation (\ref{eq:edot}) is the form for the luminosity we will use in our calculations below.

\section{Theory-Independent Equations of Motion for Compact Objects}
\label{sec:eom}

Treating compact objects as test-particles with environmentally-dependent masses, $m=m(\psi)$, gives us the following action for their world-lines:
\begin{equation}
    S=\int m(\psi) d\tau \, .
\end{equation}
Upon extremization this yields 
\begin{align}
\label{eomsens}
m \, \dot{u}^\mu 
= -\left( u^\mu u^\nu + g^{\mu \nu} \right) m' \, \psi_{, \nu} = -h^{\mu \nu} m' \, \psi_{,\nu} = -m' \, D^\mu \psi \, , 
\end{align}
where $\dot{u}^\mu = u^\nu\nabla_\nu u^\mu$ is the 4-acceleration of the particle, where the prime denotes differentiation with respect to $\psi$, where $h_{\mu \nu} = g_{\mu \nu}+u_{\mu} u_{\nu}$ is the spatial projection tensor, and where $D^\mu X = h^{\mu \nu}X,_\nu$ is the spatially-projected derivative operator. This provides the equation of motion for a sensitive body in any metric theory of gravity.

Moving on to the stress-energy tensor of these bodies, we can write
\begin{equation}
    T^{\mu \nu} (y^{\alpha})= \int \frac{m(\psi) \delta^4 (y^\alpha - x^\alpha)}{\sqrt{-g}} \frac{dx^\mu}{d\tau} \frac{dx^\nu}{d\tau} \, d\tau \,, 
    \label{eq:stress_energy_tensor1}
\end{equation}
where $\psi$ are again arbitrary external fields, $g$ is the determinant of the metric, and $\tau$ is the proper time of the particle along its world-line $x^{\mu} = x^{\mu}(\tau)$. In Appendix B we show that this leads to the conservation equation
\begin{equation}
\label{eq:ES1}
  T^{\mu \nu}_{\phantom{\mu \nu} ;\nu}(y^{\alpha}) = \int \frac{ \delta^4(y^\alpha - x^\alpha)}{\sqrt{-g}} \left(m\, \dot{u}^\mu + {m'}\, \frac{d\psi}{d\tau} \, u^\mu \right) d\tau = \psi,^\mu \frac{dT}{d \psi}\, ,
\end{equation}
where $T=T^{\mu}_{\phantom{\mu} \mu}$ is the trace of the stress-energy tensor. Generalizing to $A$ arbitrary external fields then gives our results as
\begin{equation}
\label{eq:emvio1}
    T^{\mu\nu}_{\phantom{\mu\nu} ;\nu} = \sum_A\psi_A,^{\mu} \frac{dT}{d\psi_A} \qquad {\rm and} \qquad \dot{u}^\mu=-\sum_A\frac{1}{m}\frac{d m}{d\psi_A} \, D^\mu\psi_A \,.
\end{equation}
Although these equations appear to represent a violation of stress-energy conservation, we note that this is only as a consequence of Eardley's ansatz that compact bodies can be treated as test particles with sensitive masses. In reality, no local conservation laws are violated in any metric theory of gravity. 

In what follows we will consider the external fields $\psi_A$ to be the scalar gravitational potentials in standard post-Newtonian gauge, which for point particles are\footnote{The potentials $\Phi_3$ and $\Phi_4$ vanish for point particles, and $\Phi_5$ can be made to vanish by a suitable choice of gauge.} 
\begin{eqnarray}
\label{psiA}
\hspace{-2.5cm}\{ \psi_A \}&=\{ U, \Phi_1, \Phi_2 , \Phi_6 \}
=
\left\{ \sum_a \frac{m_{{\rm G}a}}{r_a} ,\sum_a \frac{ m_{{\rm G}a} v_a^2}{r_a},
\sum_{a, b\neq a} \frac{ m_{{\rm G}a}m_{{\rm G}b}}{r_a r_{ab}},
\sum_{a} \frac{ m_{{\rm G}a}}{r_a^3}
(\textbf{v}_a\cdot \textbf{r}_{a})^2 
\right\}
\end{eqnarray}
where $r_a=\vert \textbf{x}-\textbf{r}_a(t) \vert$, ${r}_{ab} = \vert \textbf{r}_a(t)-\textbf{r}_b(t) \vert$, and $\textbf{v}_a=d\textbf{r}_a/dt$.
The reason for this choice is firstly that we do not know the fundamental degrees of freedom in the gravitational sector of the theory, as we are working in the theory-independent PPN approach. The second reason is that we are considering metric theories of gravity only, which means that matter fields couple only to the metric (thus enforcing the weak equivalence principle). In such a situation all gravitational degrees of freedom should be communicated to the matter through the metric only. We note that such a specification of the external fields to which the mass is sensitive requires a gauge choice, so our results are valid only for the choice of standard post-Newtonian gauge.

\subsection{A Single Sensitive Body}

At leading-order (0PN) the equation of motion in Equation (\ref{eq:emvio1}) can only be sensitive to the Newtonian gravitational potential $U$. This gives
\begin{equation}
\label{eq:Geo1}
    \dot{u}^i=\frac{du^i}{dt} -\frac{1}{2} h_{00,i} = -\frac{1}{m} \frac{dm}{dU} \, U,_{i} + \mathcal{O}(\eta^3) \, .
\end{equation}
Here $h_{00}=2\alpha U$ from the PPN test metric (\ref{ppn3}), which allows us to write down a modified Newtonian gravitational force law for the sensitive compact body:
\begin{align}
    \frac{d^2 x^i}{d t^2} = (\alpha - s_U) U,_{i} + \mathcal{O}(\eta^2) \, , \qquad {\rm where} \qquad s_U = \frac{d \ln m}{dU} \, ,
    \label{eq:1}
\end{align}
which is the sensitivity of the body with respect to the potential $U$. 

If we label the sensitive body as ``1'', and consider body ``2'' to be insensitive, then we obtain the following Newtonian approximation to the equations of motion for these two bodies: 
\begin{equation}
\label{eq:t1}
    m_{{\rm } 1} \, \textbf{a}_1= -\left( \alpha-s^{(1)}_U \right) \frac{m_{{\rm } 1} \, m_{{\rm G} 2}}{r_{12}^2} \, \textbf{n}_{12}  \qquad {\rm and} \qquad m_2 \, \textbf{a}_2=\alpha \, \frac{m_{{\rm G} 1} \, m_{{\rm } 2}}{r_{12}^2} \, \textbf{n}_{12} \, ,
\end{equation}
where $\textbf{a}_1$ and $\textbf{a}_2$ are the 3-accelerations of the two bodies, $r_{12}$ is the inter-body distance, and $s^{(1)}_U$ is the sensitivity parameter of body 1. In writing this equation we have distinguished between the inertial masses of these bodies ($m_{{\rm } i}$) and their active gravitational masses ($m_{{\rm G}i}$). 

By comparing the equations in (\ref{eq:t1}) to the Newtonian equations
\begin{equation}
    m_{{\rm } 1} \, \textbf{a}_1= -\alpha \,  \frac{m_{{\rm P} 1} \, m_{{\rm G} 2}}{r_{12}^2} \, \textbf{n}_{12}  \qquad {\rm and} \qquad m_2 \, \textbf{a}_2= \alpha\, \frac{m_{{\rm G} 1} \, m_{{\rm P} 2}}{r_{12}^2} \, \textbf{n}_{12} \, ,
\end{equation}
we can identify that the insensitive body has passive gravitational mass $m_{{\rm P} 2} = m_{{\rm } 2}$, while the sensitive body has $\alpha \, m_{{\rm P} 1} = \big( \alpha-s^{(1)}_U \big) \, m_{{\rm } 1}$. Given that body 1 is sensitive, while body 2 is not, this is as expected. To go further we can insist that active and passive gravitational mass should be equivalent\footnote{It is only within fully non-conservative theories, which violate conservation laws and do not possess Lagrangians, that active and passive masses can be different \cite{will2018theory}.}, such that $m_{{\rm G} i}=m_{{\rm P} i}$. We then have
\begin{equation}
    \alpha \,m_{{\rm G} 1}=\alpha \,m_{{\rm P} 1} = \left( \alpha-s^{(1)}_U \right) \, m_{{\rm } 1} \qquad {\rm and} \qquad \alpha \, m_{{\rm G} 2}=\alpha \, m_{{\rm P} 2} =\alpha\, m_2 \, ,
\end{equation}
which then allows us to write (\ref{eq:t1}) as
\begin{equation}
\label{1sens}
    m_{{\rm } 1} \, \textbf{a}_1= -\left( \alpha-s^{(1)}_U \right) \frac{m_{{\rm } 1} \, m_{{\rm } 2}}{r_{12}^2} \, \textbf{n}_{12}  \qquad {\rm and} \qquad m_2 \, \textbf{a}_2=\left( \alpha-s^{(1)}_U \right) \, \frac{m_{{\rm } 1} \, m_{{\rm } 2}}{r_{12}^2} \, \textbf{n}_{12} \, .
\end{equation}
This has the pleasing property that the sum of Newtonian forces on the two bodies sums to zero, $\textbf{F}=\textbf{a}_1m_1+\textbf{a}_2m_2=0$, in keeping with the global conservation of linear momentum. It also allows us to deduce the absent metric contribution to sensitivities in the equation of motion {\it without} having to solve for the metric in a theory with specific field content, or even know the field equations of the underlying theory.

\subsection{Two Sensitive Bodies}
\label{sec:2sens}

The results in (\ref{1sens}) clearly generalize to the second body being sensitive, and the first insensitive, under the exchange of labels $1 \leftrightarrow 2$. However, we also want to know what happens when {\it both} bodies 1 and 2 are sensitive. Given the additive nature of sensitivity parameters in the coupling constants in (\ref{1sens}), we suppose that this should be expected to be a general feature of sensitivities and write
\begin{equation}
\label{norder}
     m_{{\rm } 1} \, \textbf{a}_1= -\left( \alpha-s^{\{12\}}_{\rm N} \right) \frac{m_{{\rm } 1} \, m_{{\rm } 2}}{r_{12}^2} \, \textbf{n}_{12}  \qquad {\rm and} \qquad m_2 \, \textbf{a}_2=\left( \alpha-s_{\rm N}^{\{12\}} \right) \, \frac{m_{{\rm } 1} \, m_{{\rm } 2}}{r_{12}^2} \, \textbf{n}_{12} \, ,
\end{equation}
such that in the limit where the sensitivity parameter of one of the bodies vanishes, the new parameter $s_{\rm N}^{\{12\}}$ reduces to that of the other:
\begin{equation}
\label{sulim}
    \lim_{s_U^{(1)} \rightarrow 0} s_{\rm N}^{\{12\}} = s_U^{(2)} \qquad {\rm and} \qquad  \lim_{s_U^{(2)} \rightarrow 0} s_{\rm N}^{\{12\}} = s_U^{(1)} \, .
\end{equation}
This guarantees we reproduce the required results from the previous section in the case when only one body is sensitive. 

Requiring that $s_{\rm N}^{\{12\}}$ is linear in both $s_U^{(1)}$ and $s_U^{(2)}$, that it is symmetric under their interchange, and that it reduces to the values given in (\ref{sulim}) under the specified limits then tells us that
\begin{equation}
\label{su0}
    s_{\rm N}^{\{12\}} = s_U^{(1)}+s_U^{(2)}-\frac{1}{c_{\rm N}} \, s_U^{(1)} s_U^{(2)} = c_{\rm N} - \frac{1}{c_{\rm N}} \left( s^{(1)}_U -c_{\rm N} \right) \left( s^{(2)}_U -c_{\rm N} \right) \, ,
\end{equation}
where $c_{\rm N}$ is a constant that we interpret as the `critical value' of the Newtonian-level sensitivity parameters $s_U^{(1)}$ and $s_U^{(2)}$ (so named as whenever either of these parameters equals $c_{\rm N}$ the sensitivity of both bodies is entirely removed from $s_U$, which then reduces to $c_{\rm N}$). At the level of our considerations, $c_{\rm N}$ is an unspecified constant. It should, however, be expressible as a function of the parameters of whatever theory is being considered. The three terms after the first equality in (\ref{su0}) might be understood, when considering the motion of (say) body 1, as being due to (i) non-geodesic motion of body 1, (ii) a modified amplitude of gravitational field of body 2, and (iii) the non-geodesic motion of body 1 due to the modification in the gravitational field strength of body 2.

\subsection{Post-Newtonian Sensitivities}

Having obtained a theory agnostic description of compact bodies at Newtonian order, we now wish to describe the dynamics of the binary systems at post-Newtonian order. For this we must add post-Newtonian sensitivity terms to the right-hand side of Equation (\ref{eq:emvio1}), which will necessarily appear as linear additions. At post-Newtonian order we expect the mass to be sensitive to not only $U$, but also to the post-Newtonian potentials $\Phi_1,\Phi_2,$ and $\Phi_6$ defined in (\ref{psiA}). From these potentials, we define the sensitivities
\begin{equation}
    s_{\Phi_1} \equiv \frac{d \ln m}{d\Phi_1}, \quad s_{\Phi_2} \equiv \frac{d \ln m}{d\Phi_2}, \quad s_{\Phi_6} \equiv \frac{d \ln m}{d\Phi_6} \quad    {\rm and} \quad s'_{U} \equiv \frac{d^2 \ln m}{dU^2} \, .
\end{equation}
Taking body 1 to be the sole sensitive body, we find that (\ref{eq:emvio1}) can be written as 
\begin{equation} \label{eq:ppN+Sen1}
\begin{split}
   m_1 \boldsymbol{a}_1 = &-\left(\alpha -s^{(1)}_U \right)\frac{m_1 m_2}{r^2}\boldsymbol{n}+\frac{m_1 m_2}{r^2}\boldsymbol{n}\left[ \left( 2\alpha\gamma+2\beta-2s^{(1)}_U(\alpha+\gamma)+s^{(1)\prime}_U \right) \frac{m_2}{r} \right.\\ 
    & \left. + \frac{1}{2}\left(\left(\alpha-s^{(1)}_U\right)\left(4\gamma +4\alpha+\alpha_1 \right)+4\beta - 2\alpha^2 - 2 f \big( s_U^{(1)} \big)  +2s^{(1)}_{\Phi_2}\right)\frac{m_1}{r}  \right. \\
    &\left. +\frac{1}{2} \left(4\gamma+4\alpha+\alpha_1 \right)\left(\boldsymbol{v}_1 \cdot \boldsymbol{v}_2 \right) -\frac{1}{2}\left( 2\gamma+2\alpha+\alpha_2-2s^{(1)}_{\Phi_1} \right)v_2^2 -\left(\gamma +s^{(1)}_U\right) v_1^2\right. \\
    &\left. +\frac{3}{2} \left(\alpha+\alpha_2+2s^{(1)}_{\Phi_6} \right)(\boldsymbol{n}\cdot\boldsymbol{v}_2)^2  \right] +\frac{m_1 m_2}{r^2}\boldsymbol{n} \cdot \left[ \left( 2\gamma+2\alpha \right) \boldsymbol{v_1}-\left( 2\gamma+\alpha+s^{(1)}_U \right)\boldsymbol{v}_2 \right] \boldsymbol{v}_1\\
    &-\frac{1}{2}\frac{m_1 m_2}{r^2}\boldsymbol{n}\cdot \left[ \left( 4\gamma+4\alpha+\alpha_1 \right) \boldsymbol{v}_1- \left( 4\gamma+2\alpha+\alpha_1-2\alpha_2-4 s^{(1)}_{\Phi_6} \right) \boldsymbol{v}_2 \right] \boldsymbol{v}_2 \, ,
\end{split}
\end{equation}
where we have suppressed the subscripts on $r$ and $\boldsymbol{n}$ to keep things as concise as possible. 

The term $f(s_U^{(1)})$ is included in the pre-factor of the term involving the combination $m_1^2\, m_2$ to account for the (as yet unspecified) contribution of the sensitivity of body 1 to the gravitational field of body 2, as encoded in the metric component
\begin{equation} \label{feq}
g_{00} \simeq -1 +2 \alpha \, U^{(2)} - 2 \beta \, \big( U^{(2)} \big)^2 + (\alpha + 2 \gamma) \, \Phi_1^{(2)} +2 \Big( \alpha^2 -2 \beta +f\big( s_U^{(1)} \big) \Big) \, \Phi_2^{(2)} \, .
\end{equation}
The superscripts in brackets in this equation indicate the dependence on each of the masses, or the domain of support in the case of gravitational potentials, and $f(s_U^{(1)})$ is included to account for the dependence of $\Phi_2^{(2)}$ on the sensitivity of body 1 through the appearance of $U^{(1)}$ in
$$
\Phi^{(2)}_2({\bf x}) \equiv \int \frac{\rho^{*(2)}({\bf x'}) U^{(1)}({\bf x'})}{\vert {\bf x} - {\bf x'}\vert} \, d^3 {x'} \, .
$$
The equation of motion from (\ref{eq:ppN+Sen1}) can be generated from the Lagrangian
\begin{align}
\label{eq:lagrangian}
L =&  - (m_1+m_2) - \left( \frac{1}{2} m_1 v_1^2  + \frac{1}{2} m_2 v_2^2  \right) + \left( \frac{1}{8} m_1 v_{1}^{4} + \frac{1}{8} m_2 v_{2}^{4} \right) +\left(\alpha - s_U^{(1)} \right)\frac{m_{1} m_{2}}{r} 
   \nonumber\\
    &+\left( \alpha+2 \gamma +s_U^{(1)} \right) \frac{m_{1} m_{2}}{2 r}  v_{1}^{2} 
    + \left(\alpha+ 2 \gamma - 2 s_{\Phi_1}^{(1)} - 2 s_{\Phi_6}^{(1)} \right) \frac{m_{1} m_{2}}{2 r} v_{2}^{2}
  \nonumber  \\
    &-  \left( 3\alpha + 4\gamma + \alpha_1 - \alpha_2 - 2 s_{\Phi_6}^{(1)}  \right) \frac{m_{1} m_{2}}{2 r}(\mathbf{v}_{1} \cdot \mathbf{v}_{2}) - \left( \alpha + \alpha_2 +2 s_{\Phi_6}^{(1)} \right) \frac{m_{1} m_{2}}{2 r} (\mathbf{v}_{1} \cdot \mathbf{n}) (\mathbf{v}_{2} \cdot \mathbf{n})
    \nonumber \\  
    &+ \left( 2\alpha^2-2\beta -(\alpha-s^{(1)}_U)^2 -s^{(1)\prime}_U \right) \frac{m_{1} m_{2}^2}{2 r^2} + \left(\alpha^2 -2\beta + f(s_U^{(1)}) -s^{(1)}_{\Phi_2} \right) \frac{m_{1}^2 m_{2}}{2r^2} \, .
\end{align}
Deriving the dynamics of the two bodies from the Lagrangian above ensures the equality of passive and active gravitational mass, as can be verified by noting that the Newtonian-order terms in the first line reproduces the equations in (\ref{1sens}), and extends this to post-Newtonian order. We can compare this to the modified EIH Lagrangian \cite{Will_2018}, described in Appendix C, to read off the parameter values in Table \ref{tab:EIH}.

We can now impose the condition that the theories of gravity we wish to consider are without preferred-frame effects. Performing a Galilean transformation on the EIH Lagrangian shows that this restricts the modified EIH parameters to satisfy \cite{Will_2018}
\begin{equation}
\mathcal{A}_a=\mathcal{B}_{[ab]}=\mathcal{C}_{ab}=\mathcal{E}_{ab}=0 \, .
\end{equation}
The PPN parameters of such theories are given by $\alpha_1=\alpha_2=0$, which immediately results in $s_{\Phi_1}=0$ and $s_{\Phi_6}=-\frac{1}{2} s_U$. The remaining non-zero parameters are then $\mathcal{G}_{12}$, $\mathcal{B}_{(12)}$, $\mathcal{D}_{122}$ and $\mathcal{D}_{211}$, and it remains to determine the $f$ that can appear in $\mathcal{D}_{abc}$. We proceed by defining $s_{\varphi_2}\equiv s_{\Phi_2} -f + 2 \alpha s_U-s_U^2$, which for body 1 sensitive gives
\begin{align}
\mathcal{D}_{122} &= (\alpha - s^{(1)}_U)^2+2(\beta -\alpha^2) +s^{(1)\prime}_U\\
\mathcal{D}_{211} &= (\alpha - s^{(1)}_U)^2+2(\beta -\alpha^2) + s_{\varphi_2}^{(1)} \, ,
\end{align}
with corresponding expression with $1 \leftrightarrow 2$ for a sensitive body 2. We note that the first term in each of these equations (i.e. the one containing $s_U$) is equal to $\mathcal{G}_{12}^2$. This re-definition gives equality between $\mathcal{D}_{122}$ and $\mathcal{D}_{211}$ up to the post-Newtonian sensitivities $s^{(1)\prime}_U$ and $s_{\varphi_2}^{(1)}$, which we expect to be useful for constructing global conservation laws.

\begin{table}[h]
    \centering
    \begin{tabular}{|c|c|c|}
        \hline
        EIH Parameter & Body 1 Sensitive & Body 2 Sensitive $\phantom{\Big(}$ \\[5pt] \hline
        $\mathcal{G}_{12}$ & $\alpha - s^{(1)}_U$ & $\alpha - s^{(2)}_U \phantom{\Big(}$\\[5pt] \hline
        $\mathcal{A}_{1}/\mathcal{A}_{2}$ & 0 & 0 \phantom{\Big(} \\[5pt] \hline
        $\mathcal{B}_{12}$ & $\frac{1}{3}\left(2\gamma+\alpha+s^{(1)}_U\right)$ & $\frac{1}{3}\left(2\gamma+\alpha-2s^{(2)}_{\Phi_1}-2s^{(2)}_{\Phi_6}\right) \phantom{\Big(}$ \\ \hline
        $\mathcal{B}_{21}$ & $\frac{1}{3}\left(2\gamma+  \alpha-2s^{(1)}_{\Phi_1}-2s^{(1)}_{\Phi_6} \right) \phantom{\Big(}$ & $\frac{1}{3}\left(2\gamma+\alpha+s^{(2)}_U\right) \phantom{\Big(}$  \\ \hline
        $\mathcal{C}_{12}$ & $\alpha_1-\alpha_2+2s^{(1)}_{\Phi_1}$ & $\alpha_1-\alpha_2+2s^{(2)}_{\Phi_1} \phantom{\Big(}$ \\[5pt] \hline
        $\mathcal{D}_{122}$ & $2(\beta -\alpha^2) +(\alpha-s^{(1)}_U)^2+s^{(1)\prime}_U $ & $2\beta - \alpha^2 -f(s_U^{(2)}) +s^{(2)}_{\Phi_2} \phantom{\Big(}$ \\[5pt] \hline
        $\mathcal{D}_{211}$ & $2\beta - \alpha^2 -f(s_U^{(1)}) +s^{(1)}_{\Phi_2}$ & $2 (\beta -\alpha^2) +(\alpha-s^{(2)}_U)^2 +s^{(2)\prime}_U \phantom{\Big(}$ \\[5pt] \hline
        $\mathcal{E}_{12}$ & $\alpha_2+2s^{(1)}_{\Phi_6}+s^{(1)}_U$ & $\alpha_2+2s^{(2)}_{\Phi_6}+s^{(2)}_U \phantom{\Big(}$  \\[5pt] \hline
    \end{tabular}
    \caption{The modified EIH parameters when one of the bodies is sensitive.}
    \label{tab:EIH}
\end{table}

The question now arises as to what happens when both bodies are sensitive. We have already devised a solution for this in the case of $\mathcal{G}_{12}$, which is given by using the expression for $s_{{\rm N}}^{\{12\}}$ in Equation (\ref{su0}), such that $\mathcal{G}_{12}=\alpha - s_{{\rm N}}^{\{12\}}$. This reduces to the appropriate expressions from Table \ref{tab:EIH} when only one body is sensitive, and is linear in the sensitivity of each body (as argued for in Section \ref{sec:2sens}). We can apply the same logic to determine that 
\begin{equation}
\mathcal{B}_{(12)}= \frac{1}{3} \left( 2 \gamma + \alpha + s_{{\rm N}}^{\{12\}} \right) \, .
\end{equation}
In writing this expression we note that the as-yet unspecified constant appearing in $s_{{\rm N}}^{\{12\}}$ is the same in both $\mathcal{G}_{12}$ and $\mathcal{B}_{(12)}$. This is a consequence of the equation of motion and the Lagrangian being defined self-consistently. 

We can now turn our attention to the remaining parameters, $\mathcal{D}_{122}$ and $\mathcal{D}_{211}$. For two sensitive bodies these are expected to be given by expressions of the form
\begin{equation} \label{d122b}
\begin{split}
    \mathcal{D}_{122}&=\left(\alpha-s^{\{12\}}_{\rm N} \right)^2 +2 (\beta -\alpha^2) +s^{(1)\prime}_U +s^{(2)}_{\varphi_2} + \dots
\\
    \mathcal{D}_{211}&= \left(\alpha-s^{\{12\}}_{\rm N} \right)^2 + 2(\beta - \alpha^2)   +s^{(2)\prime}_U +s^{(1)}_{\varphi_2}+ \dots
\, ,
\end{split}
\end{equation}
where dots denote terms that encode the response of body 1 to the change in post-Newtonian gravitational field associated with body 2 (and vice versa). For $\mathcal{D}_{122}$, we take these terms to be linear in both $s^{(1)\prime}_U$ and $s^{(2)}_{\varphi_2}$, and therefore of the form
$$
-\frac{1}{c_{\rm PN}} s^{(1)\prime}_U s^{(2)}_{\varphi_2}
=  -\frac{1}{c_{\rm PN}} \left( s^{(1)\prime}_U - c_{\rm PN} \right)\left( s^{(2)}_{\varphi_2} -c_{\rm PN}\right) + c_{\rm PN} - s^{(1)\prime}_U - s^{(2)}_{\varphi_2} \, ,
$$
with a similar expression for $\mathcal{D}_{211}$ with $1 \leftrightarrow 2$. The constant $c_{\rm PN}$ in this equation is the critical value of $s^{\prime}_U$ and $s_{\varphi_2}$, at which the sensitivity to the post-Newtonian potentials drops out. The new constant in $\mathcal{D}_{122}$ and $\mathcal{D}_{211}$ is taken to be the same in both equations to ensure they have the appropriate symmetry under the interchange $1\leftrightarrow 2$. The reader will note that we have also replaced $s_U^{(a)}$ with $s_{\rm N}^{\{12\}}$ in Equation (\ref{d122b}), to ensure they have the correct limit when only one body is sensitive, as well as appropriately encoding the response of that sensitivity body to the change in Newtonian gravitational field of the other. Our expressions for the values of the non-zero modified EIH parameters with two sensitive bodies are summarized in Table \ref{tab:EIH2}, where we have defined
\begin{equation}
s^{\{ab\}}_{{\rm PN}} \equiv c_{\rm PN} -\frac{1}{c_{\rm PN}} \big( s^{(a)\prime}_U - c_{\rm PN} \big)\left( s^{(b)}_{\varphi_2} -c_{\rm PN}\right) \, ,
\end{equation}
which is not symmetric under interchange of $a$ and $b$. 

We can use the results in Table \ref{tab:EIH2} to motivate a new set of modified PPN parameters that include the effects of sensitivity. These are given in Table \ref{tab:newppn}, and in terms of these new parameters we can write the equations of motion of our two bodies as
\begin{equation} \label{eq:ppN+Sen2}
\begin{split}
   m_1 \boldsymbol{a}_1 = &-{\tilde \alpha}\, \frac{m_1 m_2}{r^2}\boldsymbol{n}
   +\frac{m_1 m_2}{r^2}\boldsymbol{n} \cdot \left[ 2\left( \tilde{\gamma}+\tilde{\alpha} \right) \boldsymbol{v_1}-\left( 2\tilde{\gamma}+\tilde{\alpha} \right)\boldsymbol{v}_2 \right] \left( \boldsymbol{v}_1-\boldsymbol{v}_2 \right)\\[5pt]&
   +\frac{m_1 m_2}{r^2}\boldsymbol{n}\left[ 2 \left( \tilde{\alpha}\tilde{\gamma}+\tilde{\beta}_1 \right) \frac{m_2}{r}  + \left( \tilde{\alpha}\left(2 \tilde{\gamma} +\tilde{\alpha}\right)+2\tilde{\beta}_{2}\right)\frac{m_1}{r}  \right. \\
    &\left.\hspace{2.5cm} +2 \left(\tilde{\gamma}+\tilde{\alpha} \right)\left(\boldsymbol{v}_1 \cdot \boldsymbol{v}_2 \right) -\left( \tilde{\gamma}+\tilde{\alpha} \right)v_2^2 -\tilde{\gamma} \, v_1^2 +\frac{3 \tilde{\alpha}}{2}(\boldsymbol{n}\cdot\boldsymbol{v}_2)^2  \right] \, ,
\end{split}
\end{equation}
with the equation of motion for body 2 being given by $1\leftrightarrow2$. In Section \ref{sec:st} we will show that our results correctly reproduce the equations of motion for sensitive bodies in scalar-tensor theories of gravity, while in Section \ref{sec:gws} we will use them to derive a theory-independent waveform for gravitational radiation emitted from binary inspirals.

\begin{table}[t]
    \centering
    \begin{tabular}{|c|c|}
        \hline
        EIH Parameter & Form for Two Sensitive Bodies $\phantom{\Big(}$ \\[5pt] \hline
        $\mathcal{G}_{ab}$ & $\alpha - s^{\{ab\}}_{\rm N} \phantom{\Big(}$ \\[5pt] \hline
        $\mathcal{B}_{ab}$ & $\frac{1}{3}\left(2\gamma+\alpha+s^{\{ab\}}_{\rm N}\right) \phantom{\Big(}$  \\ \hline
        $\mathcal{D}_{abb}$ & $\big( \alpha - s^{\{ab\}}_{\rm N} \big)^2+2(\beta -\alpha^2) +s^{\{ab\}}_{{\rm PN}} \phantom{\Big(}$   \\[5pt] \hline
    \end{tabular}
    \caption{The non-zero modified EIH parameters for theories without preferred frame effects, when both bodies are sensitive.}
    \label{tab:EIH2}
\end{table}

\begin{table}[b]
    \centering
    \begin{tabular}{|c|c|}
        \hline
         New PPN Parameters & Definition in Terms of Old Parameters $\phantom{\Big(}$ \\ \hline
         $\tilde \alpha$ & $\alpha - s^{\{12\}}_{\rm N} \phantom{\Big(}$ \\ \hline
         $\tilde \gamma$ & $\gamma + s^{\{12\}}_{\rm N} \phantom{\Big(}$\\ \hline
         $\tilde \beta_1 - \tilde{\alpha}^2$ & $ \beta -\alpha^2 + \frac{1}{2} s^{\{12\}}_{{\rm PN}} \phantom{\Big(}$\\ \hline
         $\tilde \beta_2- \tilde{\alpha}^2$ & $ \beta -\alpha^2 + \frac{1}{2} s^{\{21\}}_{{\rm PN}} \phantom{\Big(}$ \\ \hline
         \end{tabular}
    \caption{A new set of PPN parameters that encode the sensitivities of both masses in the two-body problem.}
    \label{tab:newppn}
\end{table}

\section{Gravitational Wave Phase and Amplitude}
\label{sec:gws}

We now consider the gravitational waveform of radiation emitted from a binary coalescence. This can be split into three parts: (i) the early inspiral, (ii) the merger and (iii) the ringdown. Here we are focusing on weak-field gravity, and so will consider the inspiral only. In particular, we calculate the phase and amplitude of a gravitational waveform from the early inspiral of compact bodies in a theory-independent manner, building on the work of Refs. \cite{yunes2009fundamental,blanchet2004gravitational,yunes2009post}.

To relate the dynamics of a binary system to the gravitational wave phase, we use the equation of motion to write the binding energy of the binary as a function of the orbital angular frequency $\omega=2\pi F$. Assuming a quasi-circular orbit with no monopole radiation, this is related to the gravitational wave frequency $f=2F$. Defining the relative acceleration of the two bodies as ${a}\equiv{a}_1-{a}_2=r\omega^2$, where $r$ is the relative separation, gives 
\begin{align}
\label{eq:rela}
a  \;=\;
-\frac{M}{r^{2}} \tilde{\alpha}
-\frac{M}{r^{2}}  v^{2}\,(3\eta \tilde{\alpha} + \tilde{\gamma})
+\frac{2M^{2}}{r^{3}} \bigl(\eta \tilde{\alpha}^{2} + \tilde{\alpha}\tilde{\gamma}
- \psi \tilde{\beta}_{-} + \tilde{\beta}_{+}\bigr) \, ,
\end{align}
where $M=m_1+m_2$ is the total mass, $\eta\equiv m_1m_2/M^2$ is the dimensionless reduced mass parameter, $\tilde{\beta}_{+}\equiv \frac{1}{2}(\tilde\beta_1+\tilde\beta_2)$, $\tilde{\beta}_{-}\equiv \frac{1}{2}(\tilde\beta_1-\tilde\beta_2)$, and $\psi \equiv (m_1-m_2)/M$. We can now use (\ref{eq:lagrangian}) to calculate the binding energy of the binary in the centre of mass frame, $E=\sum_a \boldsymbol{p_a}\cdot \boldsymbol{v_a}-L$, as
\begin{align}
\label{eq:BindE}
E &=
\frac{\mu}{2}\,v^{2}
+\frac{3}{8}\mu\,(1-3\eta)\,v^{4}
-\frac{M\mu}{r}\tilde{\alpha}
\\&\quad +\frac{M\mu}{2r}v^{2}\,(\tilde{\alpha} + \eta \tilde{\alpha} + 2\tilde{\gamma})+\frac{M^{2}\mu}{2r^{2}}\,
\bigl(-\tilde{\alpha}^{2} - 2\psi \tilde{\beta}_{-} + 2\tilde{\beta}_{+}\bigr)\, , \nonumber
\end{align}
where $\mu =\eta M$. Equation (\ref{eq:rela}) can now be used to write (\ref{eq:BindE}) as a function of the gravitational wave frequency, such that
\begin{align}
\label{eq:Ef}
E= & -\frac{1}{2}  \mu \left(\pi \tilde{\alpha} Mf \right)^{2/3}
\Bigg[ 
1 + \left(\pi \tilde{\alpha} Mf \right)^{2/3}
\Bigg( 
-\frac{3}{4} - \frac{\eta}{12}
+ \frac{2 \tilde{\gamma}}{3\tilde{\alpha}} 
+ \frac{2\psi \tilde{\beta}_{-}}{3\tilde{\alpha}^{2}}
- \frac{2\tilde{\beta}_{+}}{3\tilde{\alpha}^{2}}
\Bigg)
\Bigg].
\end{align}
Differentiating Equation (\ref{eq:Ef}), and inverting to obtain $df/dE$, allows it to be combined with the flux from Equation (\ref{eq:edot}) to obtain 
\begin{equation}
    \frac{df}{dt}
    = \frac{48}{5\pi \mathcal{M}^2}u^{3}\Big[ \bar\kappa_{\rm D}  + \bar\kappa_1 \, u^{2/3}+ \bar\kappa_{\rm PN} \, u^{4/3} \Big] \, ,
\end{equation}
where we have again used Equation (\ref{eq:rela}). Here $\mathcal{M}\equiv (m_1m_2)^{3/5}(m_1+m_2)^{-1/5}$ is the chirp mass, and we have defined $u \equiv  \tilde{\alpha}\mathcal{M}\pi f$ and 
\begin{eqnarray} \nonumber
\bar{\kappa}_1 &\equiv& \frac{\kappa_1}{12}\, , \qquad  \qquad \bar{\kappa}_{\rm D} \equiv {\frac{5}{96} \kappa_{\rm D} \mathcal{G}^{2}\,\eta^{2/5}}\\
\bar\kappa_{\rm PN} &\equiv& \frac{\kappa_{1}}{3}\,\,\left(
   -\frac{3}{4}
   -\frac{\eta}{12}
   + \frac{2\tilde{\gamma}}{3\tilde{\alpha}}
   - \frac{2}{3\tilde{\alpha}^{2}}\,
     \bigl(-\psi \tilde{\beta}_{-} + \tilde{\beta}_{+}\bigr)
\right)-2\kappa_{\rm PN}. \label{eq:kpar1}
\end{eqnarray}
Note that here $\kappa_{\rm PN}$ is a re-labelling of the flux parameter $\kappa_{3}$ from Equation (\ref{eq:edot}), which combines with our new PPN parameters when the flux is written as a function of frequency in the following way:
\begin{align}
\kappa_{\rm PN} \equiv \frac{\kappa_3}{12}+\kappa_{1} \frac{   
    2 \eta-2 \tilde{\alpha} \tilde{\gamma}+4\psi \tilde{\beta}_{-} - 4\tilde{\beta}_{+} 
}{9 \tilde{\alpha}^{2}}\, .
\end{align}

We can now employ the stationary phase approximation, which derives from the asymptotic method of integration by steepest descent \cite{yunes2009post}, to write the phase as 
\cite{yunes2009fundamental,cutler1994gravitational}
\begin{equation}
\Psi(f) = -\frac{\pi}{4}-2\pi\int^{f/2}dF'\left(2-\frac{f}{F'} \right) \tau(F') \, ,
\end{equation}
where $\tau(F)\equiv {F}/\dot{F}$ and $F$ is the orbital frequency. To calculate the amplitude, it is necessary to start with the time-domain strain equation for a gravitational wave, assuming a dominant quadrupole mode (as in GR), this can be written as \cite{cutler1994gravitational} 
\begin{equation}
    h(t)= \frac{(384/5)^{1/2}\,\pi^{2/3}\,Q(\iota,\theta)\,\mu M}{D\,r(t)}
\cos\!\left(\int 2\pi f\,dt\right) =\mathcal{A}(t) \, \text{cos}\left(\phi(t)\right) \, ,
\end{equation}
where $D$ is the distance from source to detector, $r(t)$ is the time-dependent inter-body distance, and $Q(\iota,\theta)\equiv F_+\text{cos}(2\theta)(1+\text{cos}^2(\iota)/2)-F_{\times}\text{sin}(2\theta)\text{cos}(\iota)$, where $F_{+,\times}$ are the beam pattern functions and $\iota$ and $\theta$ are the inclination and polarization, respectively \cite{yunes2009fundamental}. The time-domain amplitude $\mathcal{A}(t)$ and phase $\phi(t)$ are defined by the last equality. As $\dot{\mathcal{A}}/\mathcal{A}\ll \dot{\phi}$ and $\ddot{\phi}\ll \dot{\phi}^2$, the stationary phase approximation allows the following expression for the Fourier transform of the strain in the frequency domain: 
\begin{equation}
    \widetilde{h}(f)=\frac{1}{2}\mathcal{A}(t_0)\frac{1}{\sqrt{2\dot{F}(t_0)}}e^{i\Psi}\, , 
\end{equation}
where $t_0$ is the stationary point defined by $\dot{\phi}(t_0) = \pi f$. Using these formulae, the phase and amplitude in Fourier space are 
\begin{eqnarray}
\hspace{-1cm} \Psi(f) =& -\frac{\pi}{4} -\phi_c + 2\pi ft_c+\frac{3}{128\pi}u^{-5/3}
      \left[\bar{\kappa}_1 -\frac{20}{9}u^{2/3}\bar\kappa_{\rm PN}-\frac{4}{7}\bar{\kappa}_{\rm D} u^{-2/3}\right] \, ,\\
  \hspace{-1cm}  \label{eq:amp} |\widetilde{h}(f)|=& -\left(\frac{5}{24\bar{\kappa}_1}\right)^{\frac{1}{2}}\pi^{-2/3}\frac{\mathcal{M}^{-5/6}}{D}Qf^{-7/6}\left[ 1 - \kappa_{\rm A}u^{2/3}-\frac{1}{2}\bar{\kappa}_{\rm D} u^{-2/3}\right] \, .
\end{eqnarray}
We have defined 
\begin{equation}
\kappa_{\rm A} \equiv \frac{1}{12}\,\,\left(
   -9
   +11\eta
   + \frac{4 \tilde{\gamma}}{\tilde{\alpha}}
   + \frac{32}{\tilde{\alpha}^{2}}\,
     \bigl(-\psi \tilde{\beta}_{-} + \tilde{\beta}_{+}\bigr)
\right)-3\frac{\mathcal{\kappa}_{\rm PN}}{\kappa_1} \, .
\end{equation}
We note that should our parameters be set to their respective GR values, $\tilde\alpha=1, \tilde\beta=1, \tilde\gamma=1, \bar\kappa_{\rm D}=0, \bar\kappa= 1$ and $\kappa_{\rm PN}= -1247/336-35\eta/12$, we recover the known GR results. These expressions for the gravitational wave phase and amplitude, in terms of our new PPN parameters, will allow us to constrain them with data in Section \ref{sec:constraints}. 

\section{Example: Scalar-Tensor Gravity}
\label{sec:st}

In this section we will demonstrate the ability of our approach to accommodate the gravitational wave emission from compact binary inspirals in scalar-tensor theories of gravity. Specifically, we will consider theories derived from the action
\begin{equation}
\label{eq:STaction}
   I = \frac{1}{16\pi} \int \left[ \phi R - \frac{\omega(\phi)}{\phi} g^{\mu\nu} \phi_{,\mu} \phi_{,\nu} \right] \sqrt{-g} \, d^4x - \sum_a \int m_a(\phi) \, d\tau_a \, ,
\end{equation}
where $\phi$ is a new fundamental scalar field, $\omega(\phi)$ is the coupling function between scalar and tensor gravitational fields, and where the integrals over proper time, $\tau_a$, are along the world-lines of each of the bodies, $a$. Each mass, $m_a$, represents a compact body, treated as a point particle dependent upon the external scalar field, as in Ref. \cite{eardley1975observable}.

Varying the action in Equation (\ref{eq:STaction}) with respect to the $g_{\mu\nu}$ and $\phi$ yields
\begin{equation}
\begin{aligned}
    G_{\mu\nu} &= \frac{8\pi G}{\phi} T_{\mu\nu} + \frac{\omega(\phi)}{\phi^2} \left( \phi_{,\mu} \phi_{,\nu} - \frac{1}{2}g_{\mu\nu} \phi_{,\lambda} \phi^{,\lambda} \right) + \frac{1}{\phi} (\phi_{;\mu\nu} - g_{\mu\nu} \Box_g \phi) \,, \\
    \Box_g \phi &= \frac{1}{3 + 2\omega(\phi)} \left( 8\pi GT - 16\pi G \phi \frac{\partial T}{\partial \phi} - \frac{d\omega}{d\phi} \phi_{,\lambda} \phi^{,\lambda} \right).
\end{aligned}
\end{equation}
Now, writing the scalar fields as a background value and a perturbation, via $\phi=\phi_0(1+\Psi)$, allows us to expand the mass of each body as
\begin{equation}
    m_a(\phi) = m_a \left[ 1 + s_a \Psi + \frac{1}{2} (s_a^2 + s_a' - s_a) \Psi^2 + O(\Psi^3) \right],
\end{equation}
where
\begin{equation}
    s_a \equiv \left( \frac{d \ln m_a(\phi)}{d \ln \phi} \right)_0, \qquad {\rm and} \qquad
    s'_a \equiv \left( \frac{d^2 \ln m_a(\phi)}{d (\ln \phi)^2} \right)_0 \, .
\end{equation}
This allows the equation of motion for body 1, in a two-body system, to be written \cite{mirshekari2013compact}:
\begin{equation}
\label{eq:STeom}
\begin{aligned}
    {\bf a}_1 &= -\frac{G\alpha m_2}{r^2} {\bf n} + \frac{G\alpha m_2}{r^2} (\mathbf{v}_1 - \mathbf{v}_2) \left[(4 + 2\bar{\gamma})\mathbf{v}_1 \cdot \mathbf{n} - (3 + 2\bar{\gamma})\mathbf{v}_2 \cdot \mathbf{n} \right]\\
    & \quad\; + \frac{G\alpha m_2}{r^2} {\bf n} \Bigg\{ -(1+\bar{\gamma})v_1^2 - (2+\bar{\gamma})(v_2^2 - 2\mathbf{v}_1 \cdot \mathbf{v}_2) + \frac{3}{2}(\mathbf{v}_2 \cdot \mathbf{n})^2 \\[-10pt]
    & \hspace{4cm} + \left[4 + 2\gamma + 2\bar{\beta}_1\right] \frac{G\alpha m_2}{r} + \left[5 + 2\gamma + 2\bar{\beta}_2\right] \frac{G\alpha m_1}{r} \Bigg\} \, ,
\end{aligned}
\end{equation}
with ${\bf a}_2$ given by a similar expression under the interchange $\{1 \rightleftharpoons 2, \textbf{n}\rightarrow -\textbf{n}\}$. The sensitive PPN parameters, $\{{\alpha}, \bar{\gamma},\bar{\beta}_1, \bar{\beta}_2\}$, are then  
\begin{equation}
\begin{aligned}
    {\alpha} =& 1 - \zeta + \zeta(1-2s_1)(1-2s_2)\, , \qquad \qquad \; \; \bar{\gamma} = -2\alpha^{-1}\zeta(1-2s_1)(1-2s_2)\\
    \bar{\beta}_1 =& \alpha^{-2}\zeta(1-2s_2)^2(\lambda_1(1-2s_1) + 2\zeta s'_1) \, , \quad \bar{\beta}_2 = \alpha^{-2}\zeta(1-2s_1)^2(\lambda_1(1-2s_2) + 2\zeta s'_2) \, , \nonumber
\end{aligned}
\end{equation}
where $\zeta \equiv (4 + 2\omega_0)^{-1}$, $\lambda_1 \equiv \left( d\omega/d\varphi \right)_0 \zeta^2/(1 - \zeta)$, $G \equiv \phi_0^{-1}(4+2\omega_0)/(3+2\omega_0)$, and where a subscript ``$0$'' denotes a quantity evaluated at $\phi_0$.

We can now make a comparison between our theory-independent two-body equation of motion, given in Equation (\ref{eq:ppN+Sen2}), and the scalar-tensor example above. For a single sensitive body we find, to leading order, that
\begin{equation}
 s_U^i = 2G \, \zeta \, s_i \, .
\end{equation}
This equation can also be seen to come directly from the scalar-tensor result $d \ln \phi / d U = 2 G \, \zeta$. Generalizing to two sensitive bodies, we can see that we recover Equation (\ref{su0}) with critical values $c_{\rm N}=G\zeta$. For the post-Newtonian sensitivities we find
\begin{align}
    s_U'^i = 2G^2\zeta(-\lambda_1 s_i+\zeta s'_i) \qquad {\rm and} \qquad
    s_{\varphi_2}^i = -4\zeta\lambda_1G^2(s_i+s_i^2)
\end{align}
with critical value $c_{\rm PN}=G^2\zeta\lambda_1$. This, together with the preferred-frame conditions from Section \ref{sec:eom}, are then entirely sufficient to show that the two-body equations of motion of scalar-tensor theories are indeed a limit of our theory-independent equations.

Continuing to calculate the flux within this class of scalar-tensor theories, one can carry out a multipolar post-Minkowskian expansion of the vacuum field equations and solve them in the wave zone for both the tensorial and scalar waveforms. From these, the flux can be calculated directly \cite{bernard2022gravitational}. These fluxes can then be specialised to circular orbits, and cast in terms of frequency, using an identical process to the one set out in Section \ref{sec:gws}. The resulting tensorial flux to post-Newtonian order is 
\begin{align}
\mathcal{F}_{\rm tensor}=\frac{32 x^{5} \mu^{2} \left(1 + \tfrac{1}{2}\bar{\gamma}\right)}{5 G \bar\alpha}
\left( 1 + \frac{x}{336} \left\{ -1247 - 896\bar{\beta}_{+} - 448\bar{\gamma} 
+ 896\bar{\beta}_{-}\psi - 980\eta \right\} \right),
\end{align}
and the leading contribution to the scalar flux is 
\begin{equation}
\mathcal{F}_{\text{scalar}} = \frac{{4 x^{5} \mu^{2} \zeta}}{{3 G \bar\alpha}} \frac{ \mathcal{S}_{-}^{2}}{ x}\, . 
\end{equation}
In these expressions $x\equiv (GM\bar\alpha\pi f)^{2/3}$, $\mathcal{S}_- \equiv {(s_2-s_1)}/{\sqrt{\bar\alpha}}$ , $\bar\beta_+ \equiv \frac{1}{2} (\bar\beta_1+\bar\beta_2)$ and $\bar\beta_-= \frac{1}{2} (\bar\beta_1-\bar\beta_2)$. These results can be understood, in terms of our theory-independent flux parameters, by writing
\begin{align}
\bar \kappa_1&=\frac{1}{\bar\alpha}\left(1+\frac{\bar\gamma}{2} \right), \quad
    \bar \kappa_{\rm D}= \zeta\frac{5(s_2-s_1)^2}{36G\bar\alpha^2}, \nonumber \\ 
    \kappa_{\rm PN}&=\frac{1+\bar\gamma/2}{336\bar\alpha} \left( -1247 - 896\bar{\beta}_{+} - 448\bar{\gamma} 
+ 896\bar{\beta}_{-}\psi - 980\eta \right). 
\end{align}
It should be noted that for this comparison to remain valid for flux and phase expressions we have assumed the quadrupole to be the leading contribution and the dipole to be sub-dominant. This demonstrates that our theory-independent approach can accommodate both the post-Newtonian dynamics and the gravitational wave emission from compact binaries within this classes of scalar-tensor theories. 

\section{Observational Constraints}\label{sec:constraints}

Now that we have obtained an expression for the early inspiral phase of a gravitational wave in terms of the PPN parameters and our new sensitivity parameters, we can attempt to constrain them with gravitational wave data. 

The first direct detection of gravitational waves from a compact binary system, GW150914, took place in 2015 and was used to place constraints on the possible deviations from GR in the phase  \cite{abbott2016tests}. Similar constraints have been obtained from subsequent detections \cite{abbott2019tests,sanger2024tests}. These bounds are placed on possible deviations from the GR post-Newtonian expression for the phase 
\begin{equation}
    \Psi(f)=\Psi^{\text{GR}}(f)+\delta \Psi(f) 
\end{equation}
where 
\begin{equation}
    \Psi^{\text{GR}}(f) = 2\pi f t_c - \phi_c - \frac{\pi}{4} 
    + \frac{3}{128 \eta \nu^5}  \sum_{n=0}^{7} \psi_n^{\text{GR}} \nu^n
    \label{eq:gr_waveform}
\end{equation}
is the usual form of this expression in GR, and where $\delta \Psi(f)$ are phase correction terms deriving from possible non-GR dynamics, which can be expanded as
\begin{equation}
\delta \Psi(f) = \frac{3}{128 \eta \nu^5}  \sum_{n=-2}^{7} \delta \psi_n \nu^n .
\label{eq:gr_wfcorrection}
\end{equation}
Here $\psi_n^{\rm GR}$ are the $(n/2)$PN coefficients in GR, which depend only on the intrinsic parameters of the binary, and $\delta \psi_n$ represent the deviations in the $({n}/{2})$PN coefficients. As before, $t_c$ and $\phi_c$ are the time and phase at coalescence, $\nu=(2\pi fM)^{1/3} $, $M=m_1+m_2$ is the total mass and $\eta= m_1m_2/M^2$ the symmetric mass ratio.  

Fractional deviation parameters can now be defined as 
\begin{equation}
    \delta \hat{\phi}_n \equiv \frac{\delta \psi_n}{\psi_n^{\text{GR}}}, 
    \label{eq:fracdevi}
\end{equation}
which are the quantities used to provide numerical constraints on deviations from GR by the LIGO/Virgo collaboration using Bayesian inference techniques. In particular, one-dimensional posteriors for the deviation parameters are obtained by keeping all other deviation parameters fixed and then varying over all other relevant quantities (e.g. mass, spin, extrinsic parameters). The deviation parameter that we are most interested in is $\delta \hat{\phi}_2$, which can be found by noting that $\delta \psi_2=\bar{\kappa}_{\rm PN}-\psi_2^{\rm GR}$.
The resulting expression for $\delta\hat{\phi}_2$ will, in general be a complicated function of sensitivities $s_i$, post-Newtonian parameters and radiative flux parameters $\kappa_i$. 

Simplifying to the special case where sensitivities are neglected, such that $s_i=0$, and assuming $\kappa_1$ and $\kappa_{\rm PN}$ take their GR values, we find
\begin{equation} \label{phi2simp}
\delta {\psi}_2 \simeq \frac{8}{3} \left( \frac{\gamma}{\alpha} - \frac{\beta}{\alpha^2} \right)\, .
\end{equation}
This shows the influence of the modified two-body dynamics on the phase of emitted radiation, without the inclusion of possible additional physics from sensitivities and without any terms beyond the ones from GR in the luminosity equation, (\ref{eq:edot}). This is not entirely self-consistent, as theories in which $\delta\hat{\phi}_2$ are allowed to take non-GR values should probably also be expected to result in changes to the other $\delta\hat{\phi}_i$. Nevertheless, it results in a simple expression, and follows (at least in part) the same logic as the LIGO/Virgo tests of gravity \cite{abbott2016tests,abbott2019tests,sanger2024tests}. More general cases can be studied as desired.

Constraints on $\delta\hat{\phi}_2$ have been obtained from GW150914   \cite{abbott2016tests}, GW170817\footnote{This constraint is relaxed somewhat, to $\delta\hat{\phi}_2-0.14^{+0.16}_{-0.15}$ with $90\%$ confidence, if neutron star tidal effects are included for one of the bodies \cite{abbott2019tests}.} \cite{abbott2019tests}, and GW230529 \cite{sanger2024tests}. These are, respectively, the first-ever direct detection of gravitational waves from binary black holes, the binary neutron star merger that had an optical counterpart, and a recent neutron star and low-mass-compact object merger. Numerical values for these constraints, obtained by allowing only $\delta\hat{\phi}_2$ to take non-GR values, are given in Table \ref{obstab} along with distance estimates. In presenting this data we have chosen to symmetrize the errors by taking the larger of the published bounds and adjusted them so that they cover $68\%$ of the probability space by assuming a Gaussian profile.

\begin{table}[t]
    \centering
    \begin{tabular}{|c|c|c|c|}
        \hline
      Event & Constraint on $\delta\hat{\phi}_2$ & Distance to Source/Mpc & symmetric mass, $\eta$\\[5pt]  \hline
        GW150914 & $ -0.35 \pm 0.21 $  &  $410 \pm 109$ & $0.247 \pm 0.003$ \\[5pt] \hline
        GW170817 & $ -0.05 \pm 0.03 $ & $41 \pm 2$ & $0.24 \pm 0.01$ \\[5pt] \hline
        GW230529 & $ -0.15 \pm 0.09 $ &  $201 \pm 62$ & $0.20 \pm 0.03$ \\[5pt] \hline
    \end{tabular}
    \caption{The constraints on $\delta\hat{\phi}_2$, the distance estimate in Mpc, and the symmetric mass ratio parameter $\eta$ for events GW150914 \cite{abbott2016tests}, GW170817 \cite{abbott2019tests}, and GW230529 \cite{sanger2024tests}, keeping all other $\delta\hat{\phi}_i$ at their GR values. Errors have been conservatively symmetrized, using the largest bound, and adjusted to 68\% confidence region assuming that they are Gaussian distributed. The central value of $\eta$ for GW170817 has been obtained from Ref. \cite{abbott2017gw170817} by taking the halfway point between the upper and lower bounds on the masses of the two bodies.}
    \label{obstab}
\end{table}

\begin{figure}[h]    
    \centering    
    \includegraphics[width=0.8\textwidth]{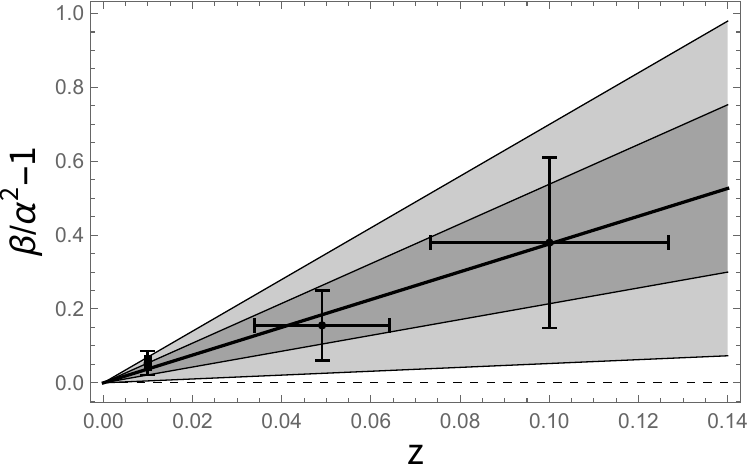} 
    \caption{Constraints on $\beta/\alpha^2-1$ as a function of cosmological redshift, for the simple relation in Equation (\ref{phi2simp}), and the data from Table \ref{obstab}. Shaded regions correspond to $68\%$ and $95\%$ confidence intervals, with linear regression having been performed to maximize a Gaussian likelihood function. We have taken $H_0\simeq 73\, {\rm km} \, {\rm s}^{-1}  {\rm Mpc}^{-1}$ \cite{riess2022comprehensive}, and have used $\gamma \simeq \alpha$ over cosmic history \cite{thomas2024constraining}.
    }  
    \label{fig3:GW plot}
\end{figure}

We can now use Equation (\ref{phi2simp}), and the data in Table \ref{obstab}, to constrain the time-evolution of the PPN parameters. In order to demonstrate the complementary nature of gravitational wave constraints to cosmological constraints, we use the result $\gamma \simeq \alpha$ from observations of the CMB \cite{thomas2024constraining}. We are then left with only the second term on the right-hand side of Equation (\ref{phi2simp}), which will give us $\beta$ as a fraction $\alpha^2$. For the redshift to the source we use the central SH0ES value of $H_0 \simeq 73 \, {\rm km} \, {\rm s}^{-1}  {\rm Mpc}^{-1}$ \cite{riess2022comprehensive}. The results of this are shown in Figure \ref{fig3:GW plot}, together with the line of best fit and the $68\%$ and $95\%$ confidence intervals, which are obtained by maximizing a Gaussian likelihood function. Errors are assumed to be uncorrelated in this simple fitting procedure, and have been combined in quadrature for the likelihood analysis.

The resultant constraint on the time-evolution of PPN parameters is
\begin{equation} \label{bcon}
\frac{\beta}{\alpha^2} \simeq 1+ \Big( 3.8 \pm 1.6 \Big) \, \times \, z \, ,
\end{equation}
where uncertainty is quoted at 1$\sigma$. This result is remarkable as it is the first ever constraint on the evolution of the PPN parameter $\beta$ over cosmic time. The result is compatible with the general relativistic value of $\beta/\alpha^2=1$ at around $2.3\, \sigma$, but one should take this result very tentatively; it has been derived under a number of very specific assumptions, and using a very simple fitting procedure. It has also been derived using only 4 data points, including the requirement that $\beta/\alpha^2=1$ at $z=0$. A more careful treatment of the data and the statistical fit is required, as well as more data, before any strong conclusions can be drawn.

\section{Discussion}
\label{sec:dis}

The constraints presented in Equation (\ref{bcon}) and Figure \ref{fig3:GW plot} are derived under the assumption of vanishing sensitivities $s_i$, and under the condition that the radiative flux parameters $\kappa_i$ are given by their GR values. In general, in cases where sensitivities are non-zero, the constraints will instead be on the dressed parameters from Table \ref{tab:newppn}, as well as involving parameters such as $\kappa_1$ and $\kappa_{\rm PN}$. In the absence of any other information on these quantities from other observables, the constraints that can be obtained on the PPN parameters is severely hampered by these additional unknown quantities.

This is readily illustrated in the example of scalar-tensor gravity, as  discussed in Section \ref{sec:st}. In this case, if both compact objects are black holes then it is expected that they take sensitivities values of $1/2$ in order to satisfy the no-hair theorems. This is equal to the critical value of this parameter in these theories, and results in the dynamics of the binary system reducing to those of GR (up to a rescaling of $\alpha$ \cite{mirshekari2013compact}). In fact, even if only one of the bodies is a black hole with sensitivity of $1/2$, then the motion is equivalent to GR up to (and including) 1.5 post-Newtonian order. Theories with this feature may limit the usefulness of binary systems containing black holes for probing the time-variation of PPN parameters. They should, however, still permit constraints on $\alpha$, though it should be noted that measurements of this parameter are degenerate with those of the chirp mass. Constraints from binary systems composed of different types of compact bodies, such as neutron stars, should also help to lift the effects of screening from critical values of sensitivity parameters, as in general one would not expect such objects to have the same value of sensitivity as black holes (unless, for some reason, all compact bodies take a critical value of sensitivity, which would severely damage any efforts to constrain PPN parameters with gravitational wave detections).

The issue of the degeneracy of parameters would appear to require other observations in order to be lifted. Indeed, this is exactly what was done in producing the constraints in Section \ref{sec:constraints}, when we used the result $\alpha \simeq \gamma$ from cosmology \cite{thomas2024constraining}. It seems conceivable that information on compact bodies from other areas of astrophysics could be used to gain independent information on sensitivities, and that amplitude of gravitational wave signals or optical observations of inspiraling binary systems (such as binary pulsars) could be used for the radiative flux parameters. This would undoubtedly be a very challenging task, but would appear to be necessary in order to disentangle the parameters that are otherwise degenerate in the gravitational wave phase.

We can now compare our results to those of Sampson {\it et al.} in Ref. \cite{sampson2013rosetta}. Their work derived the equation of motion and binding energy of binary systems under the assumption that both bodies can be represented by insensitive point particles. They included the PPN parameters $\alpha_1$ and $\alpha_2$ in their analysis, but not the value of $\alpha$. They also assumed that the energy flux of gravitational waves is given by the general relativistic result, without parametrization. This resulted in a gravitational wave phase dependent on $\beta$, $\alpha_1$ and $\alpha_2$ only. Our results differ from these by the inclusion of sensitivities, as well as radiative flux parameters, and a dependence on the PPN parameter $\gamma$, which is not found in the results published in Ref. \cite{sampson2013rosetta}.

\section{Concluding remarks}
\label{sec:conc}

We have developed a theory-independent expression for the gravitational wave phase and amplitude in terms of post-Newtonian parameters, as well as sensitivities and radiative flux parameters. Within this approach, we have constructed a new set of sensitivity parameters that allow a body's mass to depend explicitly upon Newtonian and post-Newtonian potentials. This generalizes the concept of sensitivity away from domain of scalar-tensor theories, where it was originally conceived, and into the wider framework of theory-independent parametrizations of gravity. The result is a new formalism that includes the effects of modifications to gravitational coupling parameters (as accounted for by the PPN parameters), as well as body-dependent modifications (accounted for by our new sensitivity parameters), and modifications to the luminosity of binary systems (accounted for by the radiative flux parameters).

Under a certain set of assumptions, our expression for the gravitational wave phase can be used to constrain the time-evolution of the PPN parameter $\beta$ as a fraction of the square of the amplitude of the Newtonian gravitational constant $\alpha^2$. This constraint is obtained over a redshift range $z \in [0,0.1)$, and is the first ever constraint on the cosmological variation of the $\beta$ parameter, as this parameter does not have any effect at either the level of the cosmological background or at leading-order in cosmological perturbation theory. This complements constraints on the time-variation of $\alpha$ and $\gamma$ that have recently been obtained from observations of the cosmic microwave background \cite{thomas2024constraining}, and provides a new way in which gravitational wave data can be used to constrain the gravitational interaction.

Our approach is limited by its focus on fully conservative theories of gravity, which have Lagrangians valid in all inertial frames, and therefore does not cover semi-conservative theories such as Einstein-Aether or other vector-tensor theories. Consequently, a natural extension would be to permit mass dependence on further fields in order to accommodate them. Secondly, our expression for the gravitational wave flux is parametrised phenomenologically, introducing further parameters associated with luminosity into our final expressions for the phase and amplitude. This deficit could potentially be remedied by attempting to understand the flux within a more physically motivated scheme, such that the flux parameters could be related to, or understood in terms of, the parameters that appear in the equation of motion of the massive bodies. This would likely involve a parametric formulation of the post-Minkowski approach, and is left as a long-term goal. Further data or more varied binary systems would, of course, also allow tighter and more robust constraints.

Finally, we note that we have not discussed the propagation of gravitational waves between the source and observer. In some theories of gravity this aspect of the physics also differs from general relativity, often by the inclusion of a friction term involving $\delta$, such that the amplitude of the wave, $h$, satisfies \cite{friction}
$$
h'' + 2 \mathcal{H} (1-\delta) h' + k^2 h=0 \, ,
$$
where primes denotes derivatives with respect to conformal cosmological time, $\mathcal{H}$ is the conformal Hubble rate, and $k$ is the wavenumber. A non-zero $\delta$ results in a damping of the amplitude of the wave during transit, which can be understood as a modification to the luminosity distance of the source. This affect is an important part of current attempts to constrain gravity with gravitational wave signals (see e.g.\cite{chen}), but does not affect phase of the gravitational wave signal.

\section*{Acknowledgements}
Calculations were performed with the aid of Mathematica packages xAct and xPPN \cite{hohmann2021xppn}. We acknowledge support from STFC under project reference 2897578.

\section*{Appendix A: The PPN Test Metric with $\alpha$}

The PPN test metric does not usually include $\alpha$ explicitly in the leading-order perturbation to $g_{00}$. This is due to the fact that in most applications of this framework the value of such an additional parameter can be set to unity by an appropriate choice of Newton's constant, $G$. Such a re-scaling assumes that the value of Newton's constant in the system under investigation is the same as the terrestrial value, and is a perfectly reasonable one to make for gravitational physics within the Milky Way. However, it needs to be re-assessed for the binary systems that emit the gravitational waves observed by LIGO/Virgo. These sources can be at cosmologically interesting redshifts, and in such cases the evolution of PPN parameters over cosmological time needs to be taken into account (see e.g. Refs. \cite{sanghai2017parameterized, clifton2019parametrizing, anton2022momentum, thomas2023scale, thomas2024constraining}).

In terms of the value of $\alpha$, in the leading-order perturbation of $g_{00}$ in Equation (\ref{ppn1}), this means that we are at liberty to set $\alpha=1$ at the current time, but that if $\alpha$ has any dependence on cosmological time then we cannot insist that it takes the same value in systems at high redshift. This is equivalent to the observation that in alternative theories of gravity it can often be the case that Newton's constant, $G$, takes different values at different times in the history of the Universe, and that it is not always possible to choose units so that $G=1$ at all times (the Brans-Dicke theory, formulated in the Jordan frame, would be an example of this). In such cases we need to re-instate $\alpha$ explicitly in the test metric, and if we do this in the Newtonian contribution to $g_{00}$, then we have to do it at post-Newtonian orders in the metric too in order for the PPN parameters to maintain their meaning in terms of global conservation laws.

The starting point for the calculation of these conservation laws is the definition of energy-momentum pseudo-tensor $\tau^{\mu\nu}$, which is chosen to be of the form \cite{will2018theory}
\begin{equation}
\tau^{\mu\nu} \equiv (1-aU)(t^{\mu\nu}+T^{\mu\nu}) \, ,
\end{equation}
where $T^{\mu\nu}$ is the energy-momentum tensor of matter, $a$ is an as-yet unspecified constant,
and $t^{\mu\nu}$ is the contribution to $\tau^{\mu\nu}$ coming from the gravitational fields. Global conservation laws are then obtained by requiring $\tau^{\mu\nu}_{\phantom{\mu\nu},\nu}=0$, which in addition to $T^{\mu\nu}_{\phantom{\mu \nu};\nu}=0$ (neglecting sensitivities), leads to 
\begin{equation}
\label{eq:cons}
t^{\mu\nu}_{,\nu} -aU_{,\nu} t^{\mu\nu} = \Gamma^{\mu}_{\nu\lambda}T^{\nu\lambda}+\Gamma^{\lambda}_{\nu\lambda}T^{\mu\nu}+aU_{,\nu}T^{\mu\nu} \, .
\end{equation}
This equation must be integrable in order for conservation laws to exist, meaning every term must be expressible as a total derivative. 

The non-integrable terms in Equation (\ref{eq:cons}) take their simplest form if we combine PPN parameters in the metric such that it can be expressed as 
\begin{equation} \label{newg}
\begin{split}
    g_{00}=&-1+2\alpha U + 2(\psi-\beta U^2), \qquad g_{ij}= (1+2\gamma U)\delta_{ij}\\
    g_{0j}=& -\left[2(\alpha+\gamma)+\frac{1}{2}\alpha_1 \right] V_j-\frac{1}{2}\left[ \alpha+\alpha_2-\zeta_1+2\xi\right]X_{,0j},
\end{split}
\end{equation}
where
\begin{equation}
\begin{split}
\psi =&  \frac{1}{2}(2\gamma+\alpha+\alpha_3+\zeta_1-2\xi)\Phi_1-(2\beta-\alpha^2-\zeta_2-\xi)\Phi_2+(\alpha+\zeta_3)\Phi_3\\
&+(3\gamma+3\zeta_4-2\xi)\Phi_4-\frac{1}{2}(\zeta_1-2\xi)\Phi_6-\xi\Phi_W \, ,
\end{split}
\end{equation}
where gravitational potentials take their usual meanings, and where we can see that it reduces to its usual form in the limit $\alpha \rightarrow 1$ \cite{will2018theory}. With the metric written in this way the stress-energy tensor is given by
\begin{equation} \nonumber
\begin{split}
    T^{00} = & \rho^*\left[1+\Pi+\frac{1}{2}v^2+(2\alpha-3\gamma)U \right], \qquad
    T^{0j}= \rho^*v^j\left[1+\Pi+\frac{1}{2}v^2+(2\alpha-3\gamma)U \right] ,\\
    T^{ij} = &\rho^*v^iv^j\left[1+\Pi+\frac{1}{2}v^2+(2\alpha-3\gamma)U \right] +p\delta^{ij}(1+2\gamma U).
\end{split}
\end{equation}
With repeated use of the identities
\begin{equation} \nonumber
    \frac{\partial}{\partial t}(U\nabla^2U+|\mathbf{\nabla}U|)+\frac{\partial}{\partial x^j}(U\nabla^2V_j-U_{,0}U_{,j}-2U_{,k}V_{[k,j]}) = 0
\end{equation}
and 
\begin{equation} \nonumber
\label{eq:fidentity}
    4\pi\rho^*f_{,j} = -2\frac{\partial}{\partial x^k}\Gamma_{jk}(f)+U_{,j}\nabla^2 f
\end{equation}
where $\Gamma_{jk} (f) \equiv U_{,(j}f_{,k)}-\frac{1}{2}\delta_{jk}\nabla U\cdot \nabla f$ for any function $f$, we can then identify the non-integrable part of Equation (\ref{eq:cons}) as
\begin{equation}
Q^j=U_{,j}\Bigg[\frac{1}{2}(\alpha_3+\zeta_1)\rho^*v^2 +\frac{1}{8\pi}\zeta_1\nabla^2\Phi_6+\frac{1}{8\pi}\zeta_2|\mathbf{\nabla}U|^2+\zeta_3\rho^*\Pi+3\zeta_4p+\alpha_3\rho^*\mathbf{w\cdot v} \Bigg].
\end{equation}
This equation is identical to the case in which $\alpha = 1$ \cite{will2018theory}, which means that the parameters $\alpha_3$, $\zeta_1$, $\zeta_2$, $\zeta_3$ and $\zeta_4$ retain the exact same meaning after the addition of $\alpha$ to the test metric. We note that a similar argument must be applied to the parameters $\alpha_1$ and $\alpha_2$, which mediate conservation of angular momentum: if they are non-zero $t^{[0j]}$ and $t^{[jk]}$ are non-zero and this leads to non-conservation of the angular momentum tensor \cite{will2018theory}.

When specialized to point-particles in semi-conservative theories, the metric in (\ref{newg}) reduces to the one given in Equations (\ref{ppn1})-(\ref{ppn3}), which was our purpose here. We will further study global conservation laws in the presence of sensitivities in future work.

\section*{Appendix B: Equations of Conservation and Motion}

The equations of energy-momentum conservation, and the equation of motion of sensitive bodies, can be constructed in a way that is independent from the theory of gravity. To shows this, we can consider the energy momentum-tensor of a point particle $m=m(\psi_A)$, as given in Equation (\ref{eq:stress_energy_tensor1}). Taking a partial derivative gives
\begin{align*}
    T^{\mu \nu},_{\nu} &= \int m \left( \frac{\partial}{\partial y^{\nu}} \delta(y^{\alpha} - x^{\alpha}) \right) \frac{dx^\mu}{d\tau} \frac{dx^\nu}{d\tau} \frac{d\tau}{\sqrt{-g}} 
     = - \int \frac{m \, u^\mu}{\sqrt{-g}} \frac{d}{d\tau} \delta^4 (y^\alpha - x^\alpha)\,  d\tau \, ,
\end{align*}
where we have used $u^{\nu} = dx^{\nu}/d\tau$. Integrating by parts then gives
\begin{align}
\nonumber
 T^{\mu \nu},_{\nu}   = \int m' \, \psi_{, \tau} \frac{\delta^4 (y^x - x^x)}{\sqrt{-g}} u^\mu d\tau
    + \int m \, \delta^4 (y^\alpha - x^\alpha) \frac{d}{d\tau} \left( \frac{u^\mu}{\sqrt{-g}} \right) d\tau,
\end{align}
where $m'={d m}/{d \psi}$. Using
$
(\sqrt{-g}),_{\lambda} 
=\sqrt{-g}\, \Gamma^\nu_{\nu \lambda}
$
and
$\dot{u}^\mu=u^\nu\nabla_\nu u^\mu$ then gives the first equality in Equation (\ref{eq:ES1}) from Section \ref{sec:eom}. To go further, we use the equation of motion (\ref{eomsens}), which gives
\begin{align}
T^{\mu \nu}_{\phantom{\mu\nu} ;\nu} &= \int \frac{\delta^4(y^\alpha - x^\alpha)}{\sqrt{-g}} m' \left( -D^\mu U + U_{,\tau} u^\mu \right) d\tau = -\int \frac{\delta^4(y^\alpha - x^\alpha)}{\sqrt{-g}} m' \, U,^\mu d\tau \, .
\end{align}
On using $m'U,^{\mu}=dm/d x_{\mu}$ and $u_{\mu} u^{\mu} = -1$, and moving $d/dU$ outside of the integral, we then get the final equality in Equation (\ref{eq:ES1}).

\section*{Appendix C: The Modified EIH Formalism}

The modified EIH formalism is best understood through its n-body Lagrangian \cite{Will_2018}:
\begin{equation}
\begin{aligned}
    L_{\text{EIH}} = & - \sum_{a} m_{a} \left[ 1 - \frac{1}{2} v_{a}^{2} - \frac{1}{8} (1 + \mathcal{A}_{a}) v_{a}^{4} \right] \\
    & + \frac{1}{2} \sum_{a} \sum_{b \neq a} \frac{m_{a} m_{b}}{r_{ab}} \left[ \mathcal{G}_{ab} + 3 \mathcal{B}_{ab} v_{a}^{2} - \frac{1}{2}(\mathcal{G}_{ab} + 6 \mathcal{B}_{(ab)} + \mathcal{C}_{ab}) \mathbf{v}_{a} \cdot \mathbf{v}_{b} \right. \\
    & \left. - \frac{1}{2} (\mathcal{G}_{ab} + \mathcal{E}_{ab}) (\mathbf{v}_{a} \cdot \mathbf{n}_{ab}) (\mathbf{v}_{b} \cdot \mathbf{n}_{ab}) \right]  - \frac{1}{2} \sum_{a} \sum_{b \neq a} \sum_{c \neq a} \mathcal{D}_{abc} \frac{m_{a} m_{b} m_{c}}{r_{ab} r_{ac}}.
\end{aligned}
\end{equation}
The original EIH Lagrangian was developed by Einstein, Infeld and Hoffman to describe a system of gravitating point masses to post-Newtonian accuracy; in the modified EIH formalism, that same Lagrangian has been parametrised by $\{\mathcal{A}_{a},\mathcal{G}_{ab},\mathcal{B}_{ab}, \mathcal{C}_{ab},\mathcal{E}_{ab},\mathcal{D}_{abc}\}$, which should be understood as functions of the parameters of a specific theory and the structure of bodies within that theory. These parameters possess symmetry properties, which can be derived from examine each term in the Lagrangian and identifying its symmetric under exchange of bodies. Assuming passive and active gravitational masses of a body are the same then gives 
\begin{equation}
    \mathcal{G}_{ab} = \mathcal{G}_{(ab)}, \quad \mathcal{C}_{ab} = \mathcal{C}_{(ab)}, \quad \mathcal{E}_{ab} = \mathcal{E}_{(ab)}, \quad \mathcal{D}_{abc} = \mathcal{D}_{a(bc)}.
\end{equation}
One may note that $\mathcal{B}_{ab}$ has no special symmetry in general, except for cases of theories with preferred frame effects, in which case $\mathcal{B}_{ab} \neq \mathcal{B}_{ba}$ as $v_a^2 \neq v^2_b$ under the interchange $a\leftrightarrow b$. For theories without preferred frame effects one has the following additional conditions:
\begin{equation}
    \mathcal{A}_a = \mathcal{B}_{[ab]} = \mathcal{C}_{ab} =\mathcal{E}_{ab} = 0.
    \label{eq:LIrequirements}
\end{equation}

\section*{References}

\bibliographystyle{ieeetr}

\begin{thebibliography}{10}

\bibitem{will2014confrontation}
C.~M. Will, ``The confrontation between general relativity and experiment,'' {\em Living Reviews in Relativity}, vol.~17, pp.~1--117, 2014.

\bibitem{will2018theory}
C.~M. Will, {\em Theory and Experiment in Gravitational Physics}.
\newblock Cambridge University Press, 2018.

\bibitem{clifton2012modified}
T.~Clifton, P.~G. Ferreira, A.~Padilla, and C.~Skordis, ``Modified gravity and cosmology,'' {\em Physics Reports}, vol.~513, no.~1--3, pp.~1--189, 2012.

\bibitem{uzan2011varying}
J.-P. Uzan, ``Varying constants, gravitation and cosmology,'' {\em Living Reviews in Relativity}, vol.~14, no.~1, pp.~1--155, 2011.

\bibitem{sanghai2017parameterized}
V.~A. Sanghai and T.~Clifton, ``Parameterized post-Newtonian cosmology,'' {\em Classical and Quantum Gravity}, vol.~34, no.~6, p.~065003, 2017.

\bibitem{clifton2019parametrizing}
T.~Clifton and V.~A. Sanghai, ``Parametrizing theories of gravity on large and small scales in cosmology,'' {\em Physical Review Letters}, vol.~122, no.~1, p.~011301, 2019.

\bibitem{anton2022momentum}
T.~Anton and T.~Clifton, ``The momentum constraint equation in parametrized post-Newtonian cosmology,'' {\em Classical and Quantum Gravity}, vol.~39, no.~9, p.~095005, 2022.

\bibitem{thomas2023scale}
D.~B. Thomas, T.~Clifton, and T.~Anton, ``Scale-dependent gravitational couplings in parametrized post-Newtonian cosmology,'' {\em Journal of Cosmology and Astroparticle Physics}, vol.~2023, no.~4, p.~016, 2023.

\bibitem{thomas2024constraining}
D.~B. Thomas, T.~Anton, T.~Clifton, and P.~Bull, ``Constraining post-Newtonian parameters with the cosmic microwave background,'' {\em Journal of Cosmology and Astroparticle Physics}, vol.~2024, no.~9, p.~039, 2024.

\bibitem{eardley1975observable}
D.~M. Eardley, ``Observable effects of a scalar gravitational field in a binary pulsar,'' {\em Astrophysical Journal}, vol.~196, pp.~L59--L62, 1975.

\bibitem{yunes2009fundamental}
N.~Yunes and F.~Pretorius, ``Fundamental theoretical bias in gravitational wave astrophysics and the parametrized post-Einsteinian framework,'' {\em Physical Review D}, vol.~80, no.~12, p.~122003, 2009.

\bibitem{krishnendu2021testing}
N.~Krishnendu and F.~Ohme, ``Testing general relativity with gravitational waves: An overview,'' {\em Universe}, vol.~7, no.~12, p.~497, 2021.

\bibitem{sampson2013rosetta}
L.~Sampson, N.~Yunes, and N.~Cornish, ``Rosetta stone for parametrized tests of gravity,'' {\em Physical Review D}, vol.~88, no.~6, p.~064056, 2013.

\bibitem{goldberg2017cosmology}
S.~R. Goldberg, T.~Clifton, and K.~A. Malik, ``Cosmology on all scales: A two-parameter perturbation expansion,'' {\em Physical Review D}, vol.~95, no.~4, p.~043503, 2017.

\bibitem{will1972conservation}
C.~M. Will and K.~Nordtvedt~Jr, ``Conservation laws and preferred frames in relativistic gravity,'' tech. rep., Cornell University, 1972.

\bibitem{lee1974conservation}
D.~L. Lee, A.~P. Lightman, and W.-T. Ni, ``Conservation laws and variational principles in metric theories of gravity,'' {\em Physical Review D}, vol.~10, no.~6, p.~1685, 1974.

\bibitem{nordtvedt1968equivalence}
K.~Nordtvedt~Jr, ``Equivalence principle for massive bodies. II. Theory,'' {\em Physical Review}, vol.~169, no.~5, p.~1017, 1968.

\bibitem{Taherasghari:2022wfs}
F.~Taherasghari and C.~M. Will, ``Modified geodesic equations of motion for compact bodies in alternative theories of gravity,'' {\em Physical Review D}, vol.~106, no.~6, p.~064021, 2022.

\bibitem{foster2007strong}
B.~Z. Foster, ``Strong field effects on binary systems in Einstein-{\AE}ther theory,'' {\em Physical Review D}, vol.~76, no.~8, p.~084033, 2007.

\bibitem{yagi2014constraints}
K.~Yagi, D.~Blas, E.~Barausse, and N.~Yunes, ``Constraints on Einstein-{\AE}ther theory and Ho{\v{r}}ava gravity from binary pulsar observations,'' {\em Physical Review D}, vol.~89, no.~8, p.~084067, 2014.

\bibitem{mirshekari2013compact}
S.~Mirshekari and C.~M. Will, ``Compact binary systems in scalar-tensor gravity: Equations of motion to 2.5 post-Newtonian order,'' {\em Physical Review D}, vol.~87, no.~8, p.~084070, 2013.

\bibitem{lang2014compact}
R.~N. Lang, ``Compact binary systems in scalar-tensor gravity. II. Tensor gravitational waves to second post-Newtonian order,'' {\em Physical Review D}, vol.~89, no.~8, p.~084014, 2014.

\bibitem{Will_2018}
C.~M. Will, ``Testing general relativity with compact-body orbits: A modified Einstein–Infeld–Hoffmann framework,'' {\em Classical and Quantum Gravity}, vol.~35, p.~085001, 2018.

\bibitem{thorne1980multipole}
K.~S. Thorne, ``Multipole expansions of gravitational radiation,'' {\em Reviews of Modern Physics}, vol.~52, no.~2, p.~299, 1980.

\bibitem{damour1992tensor}
T.~Damour and G.~Esposito-Farese, ``Tensor-multi-scalar theories of gravitation,'' {\em Classical and Quantum Gravity}, vol.~9, no.~9, p.~2093, 1992.

\bibitem{bernard2022gravitational}
L.~Bernard, L.~Blanchet, and D.~Trestini, ``Gravitational waves in scalar-tensor theory to one-and-a-half post-Newtonian order,'' {\em Journal of Cosmology and Astroparticle Physics}, vol.~2022, no.~8, p.~008, 2022.

\bibitem{peters1963gravitational}
P.~C. Peters and J.~Mathews, ``Gravitational radiation from point masses in a Keplerian orbit,'' {\em Physical Review}, vol.~131, no.~1, p.~435, 1963.

\bibitem{blanchet2004gravitational}
L.~Blanchet, T.~Damour, G.~Esposito-Farese, and B.~R. Iyer, ``Gravitational radiation from inspiralling compact binaries completed at the third post-Newtonian order,'' {\em Physical Review Letters}, vol.~93, no.~9, p.~091101, 2004.

\bibitem{yunes2009post}
N.~Yunes, K.~Arun, E.~Berti, and C.~M. Will, ``Post-circular expansion of eccentric binary inspirals: Fourier-domain waveforms in the stationary phase approximation,'' {\em Physical Review D}, vol.~80, no.~8, p.~084001, 2009.

\bibitem{cutler1994gravitational}
C.~Cutler and E.~E. Flanagan, ``Gravitational waves from merging compact binaries: How accurately can one extract the binary’s parameters from the inspiral waveform?,'' {\em Physical Review D}, vol.~49, no.~6, p.~2658, 1994.

\bibitem{abbott2016tests}
B.~P. Abbott {\em et~al.}, ``Tests of general relativity with GW150914,'' {\em Physical Review Letters}, vol.~116, no.~22, p.~221101, 2016.

\bibitem{abbott2019tests}
B.~P. Abbott {\em et~al.}, ``Tests of general relativity with GW170817,'' {\em Physical Review Letters}, vol.~123, no.~1, p.~011102, 2019.

\bibitem{sanger2024tests}
E.~M. S{\"a}nger {\em et~al.}, ``Tests of general relativity with GW230529: A neutron star merging with a lower–mass–gap compact object,'' {\em arXiv preprint arXiv:2406.03568}, 2024.

\bibitem{abbott2017gw170817}
B.~P. Abbott {\em et~al.}, ``GW170817: Observation of gravitational waves from a binary neutron star inspiral,'' {\em Physical Review Letters}, vol.~119, no.~16, p.~161101, 2017.

\bibitem{riess2022comprehensive}
A.~G. Riess {\em et~al.}, ``A comprehensive measurement of the local value of the Hubble constant with $1~\mathrm{km\,s^{-1}\,Mpc^{-1}}$ uncertainty from the Hubble Space Telescope and the SH0ES team,'' {\em The Astrophysical Journal Letters}, vol.~934, no.~1, p.~L7, 2022.

\bibitem{friction}
E.~Belgacem, Y.~Dirian, S.~Foffa, and M.~Maggiore, ``Gravitational-wave luminosity distance in
modified gravity theories,'' {\em Phys. Rev. D}, vol.~97, no.~10, p.~104066, 2018.

\bibitem{chen}
A.~Chen, R.~Gray, and T.~Baker, ``Testing the nature of gravitational wave propagation using dark sirens and galaxy catalogues,'' {\em Journal of Cosmology and Astroparticle Physics}, vol.~2024, no.~02, p.~035 (2024).

\bibitem{hohmann2021xppn}
M.~Hohmann, ``XPPN: An implementation of the parametrized post-Newtonian formalism using xAct for Mathematica,'' {\em The European Physical Journal C}, vol.~81, no.~6, p.~504, 2021.

\end{thebibliography}

\end{document}